\documentclass[journal=aamick,manuscript=article]{achemso}

\usepackage[dvipsnames,table]{xcolor}
\usepackage[version=3]{mhchem}
\usepackage{miller}
\usepackage{subcaption}
\usepackage{multirow}
\usepackage{color}
\usepackage{textgreek}

\usepackage{amsmath,amssymb}
\graphicspath{{Figures/}}

\usepackage{pdfpages}
\usepackage{rotating}
\usepackage[T1]{fontenc}
\usepackage[%
  breaklinks,
  bookmarks = false,
  pdfpagemode   = UseNone,%
  pdfstartview  = FitH,
  pdfstartpage  = 1,
  colorlinks    = true,
]{hyperref}

\newcommand\myshade{85}
\colorlet{mylinkcolor}{YellowOrange}
\colorlet{myurlcolor}{Aquamarine}
\colorlet{mycitecolor}{violet}

\hypersetup{
  linkcolor  = mylinkcolor!\myshade!black,
  citecolor  = mycitecolor!\myshade!black,
  urlcolor   = myurlcolor!\myshade!black,
  colorlinks = true,
}



\author{Md Salman Rabbi Limon}
\affiliation{Department of Mechanical and Aerospace Engineering, Texas Tech University, Lubbock, Texas 79409, United States}
\author{Matthew Beltran}
\affiliation{Department of Materials Science and Engineering, University of Texas at Dallas, Richardson, Texas 75080, United States}
\author{Boyu Wang}
\affiliation{Department of Materials Science and Engineering, University of Texas at Dallas, Richardson, Texas 75080, United States}
\author{Laisuo Su}
\affiliation{Department of Materials Science and Engineering, University of Texas at Dallas, Richardson, Texas 75080, United States}
\author{Zeeshan Ahmad}
\affiliation{Department of Mechanical and Aerospace Engineering, Texas Tech University, Lubbock, Texas 79409, United States}
\email{zeeahmad@ttu.edu}

\title{Space Charge Layer and Facile Halide Rearrangement Enable Fast Lithium-Ion Transport at Halide Solid Electrolyte Interfaces}


\begin{document}

\begin{abstract}

Halide solid electrolytes are promising candidates for energy-dense all-solid-state batteries because they combine high-voltage stability, facile processing, and fast Li-ion conduction. A recent study showed that physically mixing two halide solid electrolytes, \ce{Li2ZrCl6} (LZC) and \ce{Li3YCl6} (LYC), raises the Li-ion conductivity above that of either electrolyte, by 46\% and 58\% relative to the two components, suggesting that halide heterointerfaces offer a new route to faster ion transport; however, the origin of this enhancement remains unknown. Here, we combine density functional theory calculations, defect thermodynamics, machine-learning-driven molecular dynamics, and vibrational and Raman spectroscopy analysis to elucidate Li-ion transport in LZC, LYC, and their heterointerface. Using low-energy ordered supercells that represent the configurational disorder of these halides, we find that Li and Cl interstitials are the dominant charged defects in both materials. Forming an LZC|LYC interface favors the generation of interfacial Li Frenkel pairs, producing Li interstitials in LZC and Li vacancies in LYC, thereby increasing the charge-carrier concentration near the interface. By analyzing the angular orientations of Cl around the metal cations in molecular dynamics trajectories of a coherent interface, we show that the interface creates new stable Cl configurations that accommodate the framework rearrangement accompanying Li hops. Simulated vibrational power spectra and measured Raman spectra indicate lattice softening and octahedral distortion at the interface, consistent with the formation of these Cl configurations. These results support a cooperative mechanism in which space charge layer formation and a dynamically flexible interfacial halide framework enhance Li-ion conduction, providing atomistic design principles for superionic halide heterostructures and multicomponent solid electrolytes.

\end{abstract}

\section{Introduction}

All-solid-state batteries (ASSBs) are widely regarded as a transformative energy storage technology, promising higher energy and power densities and improved safety relative to conventional liquid-electrolyte lithium-ion (Li-ion) batteries.\cite{manthiram2017lithium,du2025halide,janekSolidFutureBattery2016} To be competitive, however, solid electrolytes must simultaneously offer fast Li-ion conduction,\cite{wangDesignPrinciplesSolidstate2015} chemical and electrochemical stability,\cite{kimPredictingWettabilityElectrochemical2019,xiaoUnderstandingInterfaceStability2020} mechanical compliance,\cite{ahmadStabilityElectrodepositionSolidSolid2017,tuElectrodepositionMechanicalStability2020} low cost,\cite{albertusStatusChallengesEnabling2018,kwakNewCostEffectiveHalide2021} and facile processability.\cite{tanEnablingThinFlexible2019,leePhysioElectrochemicallyDurableDryProcessed2023} Among the three major classes of inorganic solid electrolytes (oxides, sulfides, and halides), halides have recently emerged as leading candidates because they combine high ionic conductivity, good oxidative stability, and low cost.\cite{li2020progress,liang2021metal,wang2022prospects,nie2023halide,bouguern2026comparative,he2023halogen} Their compatibility with high-voltage oxide cathodes and their mechanical softness, which promotes intimate cathode--electrolyte contact at modest processing temperatures, make them particularly attractive for practical ASSBs.\cite{asanoSolidHalideElectrolytes2018,kwakNewCostEffectiveHalide2021,molaiyan2025chemistry,kwak2022emerging,du2025halide,li2025building,tuo2023recent}

The most widely studied halide electrolytes belong to the \ce{Li3MX6} family (M = Y, In, Sc, Er, Ho; X = Cl, Br, I), in which ionic transport, electrochemical stability, and mechanical properties can be tuned through the choice of metal cation and halide anion, the Li-vacancy concentration, aliovalent substitution, site occupancy, and local structural disorder.\cite{kim2021material,liang2020site,wang2023new,wang2024designing,liuStructureLi3YCl6Its2026} More recently, \ce{Li2ZrCl6} (LZC), which contains a tetravalent metal cation, has emerged as a low-cost alternative free of rare-earth elements;\cite{kwakNewCostEffectiveHalide2021,lee2024synergistic} its conductivity can be further enhanced by aliovalent doping with \ce{Fe^3+} or \ce{Ga^3+} and by engineering Zr disorder.\cite{zhongEffectCationDisorder2024,wuEngineeringLi2ZrCl6Ga32026} The compositional landscape has since expanded to disordered halospinels, oxygen-tuned and amorphous aluminum oxychlorides, and oxyhalide/garnet hetero-electrolytes,\cite{zhou2020new,kim2025oxygen,huang2026amorphous,palmer2026conformal} while wet-chemistry and liquid-phase routes enable scalable synthesis with controlled microstructure.\cite{li2019water,wang2021universal,gamo2025liquid,kwak2022emerging,li2025building} At the cell level, performance further depends on cathode|electrolyte reactivity, mixed ionic--electronic percolation in composite cathodes, and environmental stability.\cite{kochetkov2022different,zhang2025solid,tao2023halide,tan2025interfacial,li2024unraveling,fu2025cost} With few exceptions,\cite{palmer2026conformal} however, strategies to enhance ion transport have targeted the bulk chemistry of a single compound.

An alternative strategy is to exploit interfaces between ionic conductors. Heterointerfaces have long been known to enhance ionic conductivity through space charge effects, as in \ce{LiI}--\ce{Al2O3} composites and \ce{CaF2}/\ce{BaF2} heterolayers.\cite{liangConductionCharacteristicsLithium1973,sataMesoscopicFastIon2000,maierIonicConductionSpace1995,limonGrainBoundariesCeramic2026,ahmadUnifiedConsistentElectrical2025} Recently, physically mixing two halide electrolytes, LZC and \ce{Li3YCl6} (LYC), was found to raise the ionic conductivity from 0.28~mS~cm$^{-1}$ (LZC) and 0.26~mS~cm$^{-1}$ (LYC) to 0.41~mS~cm$^{-1}$ (i.e., a 46\% enhancement) without forming a new phase, and the enhancement was even larger when the mixture was paired with cathodes (67\% enhancement).\cite{wang1+122025} Unlike classical composites, in which one phase is an ionic insulator, both components here are fast Li-ion conductors with partially occupied Li sublattices, and the microscopic origin of the enhancement remains unresolved.

Several features of LZC and LYC suggest how their interface might enhance transport. Both consist of close-packed Cl sublattices in which Li, the metal cation, and vacancies share octahedral sites. Because both phases are built on close-packed Cl frameworks, a coherent interface may form, and the mismatch in local bonding may distort the octahedra near the interface, reshaping the potential energy surface for Li migration. In addition, the difference in Li chemical potential between the two phases may drive Li redistribution across the interface, forming space charge layers that modulate the local carrier concentration.\cite{maierIonicConductionSpace1995} These considerations raise three questions: (i) Is the LZC|LYC interface structurally coherent and energetically favorable? (ii) How does the asymmetry in defect energetics redistribute Li and other charged defects to form space charge layers? (iii) How do the local interfacial structure and octahedral distortion modulate Li migration?

Answering these questions requires atomistic resolution over length and time scales sufficient to capture both defect thermodynamics and ion dynamics, which is challenging for two reasons. First, because Li sites in these halides are only partially occupied, Li vacancies and interstitials are not uniquely defined with respect to an ideal lattice, complicating the classical treatment of space charge layers. Resolving this requires a model large enough to sample configurational disorder. Second, interfacial transport requires simulation cells and time scales beyond the reach of \textit{ab initio} molecular dynamics (MD). Machine-learning interatomic potentials (MLIPs)\cite{batznerE3EquivariantGraphNeural2022,marufLearningLongRangeInteractions2025,deringerMachineLearningInteratomic2019} now enable MD with near-\textit{ab initio} accuracy at these scales, making it feasible to quantify interfacial diffusivity, transport anisotropy, and structure--transport relationships.\cite{seth2025investigating}

Here, we use the LZC|LYC heterointerface as a model system to elucidate how defect thermodynamics, space charge formation, and interfacial structure govern Li-ion transport in multicomponent halide electrolytes. We combine density functional theory (DFT) calculations, defect thermodynamics, MLIP-driven MD, vibrational power spectrum analysis, and Raman spectroscopy of LZC, LYC, and their mixture. To apply defect thermodynamics to these disordered materials, we use low-energy ordered supercells that reproduce the experimental site occupancies as pristine reference structures. Our defect formation energy calculations reveal that Li interstitials prefer sites away from the metal cations, giving rise to a degree of ordering among defects. We find that Li and Cl interstitials are the dominant donor and acceptor defects, respectively, in both LZC and LYC. By enumerating defect exchange reactions at the LZC|LYC interface at the charge-neutrality Fermi level, we identify the formation of interfacial Li Frenkel pairs as the most favorable reaction; it increases the Li interstitial concentration in LZC and the Li vacancy concentration in LYC, thereby enhancing Li-ion conductivity near the interface.

We then construct a coherent, minimal-strain LZC|LYC interface model and show that the Cl framework dynamically rearranges in response to Li migration, facilitating conduction and establishing robust percolation pathways. Distorted \ce{ZrCl6}--\ce{YCl6} octahedra at the interface provide a more flexible Cl environment than either bulk phase, creating new stable Cl configurations that promote more frequent Li hop attempts. Consistent with this picture, simulated vibrational spectra show a shift of the Li and Cl modes to lower frequencies at the interface, characteristic of lattice softening, and measured Raman spectra of the mixture indicate octahedral distortion. Together, these findings identify the origins of the enhanced conductivity of mixed-halide electrolytes and provide design principles for achieving conductivities beyond those attainable through bulk transport alone.

\section{Results and Discussion}

\subsection{Crystal structures of halide solid electrolytes}

\begin{figure}[htbp]
    \centering
    \includegraphics[width=\textwidth]{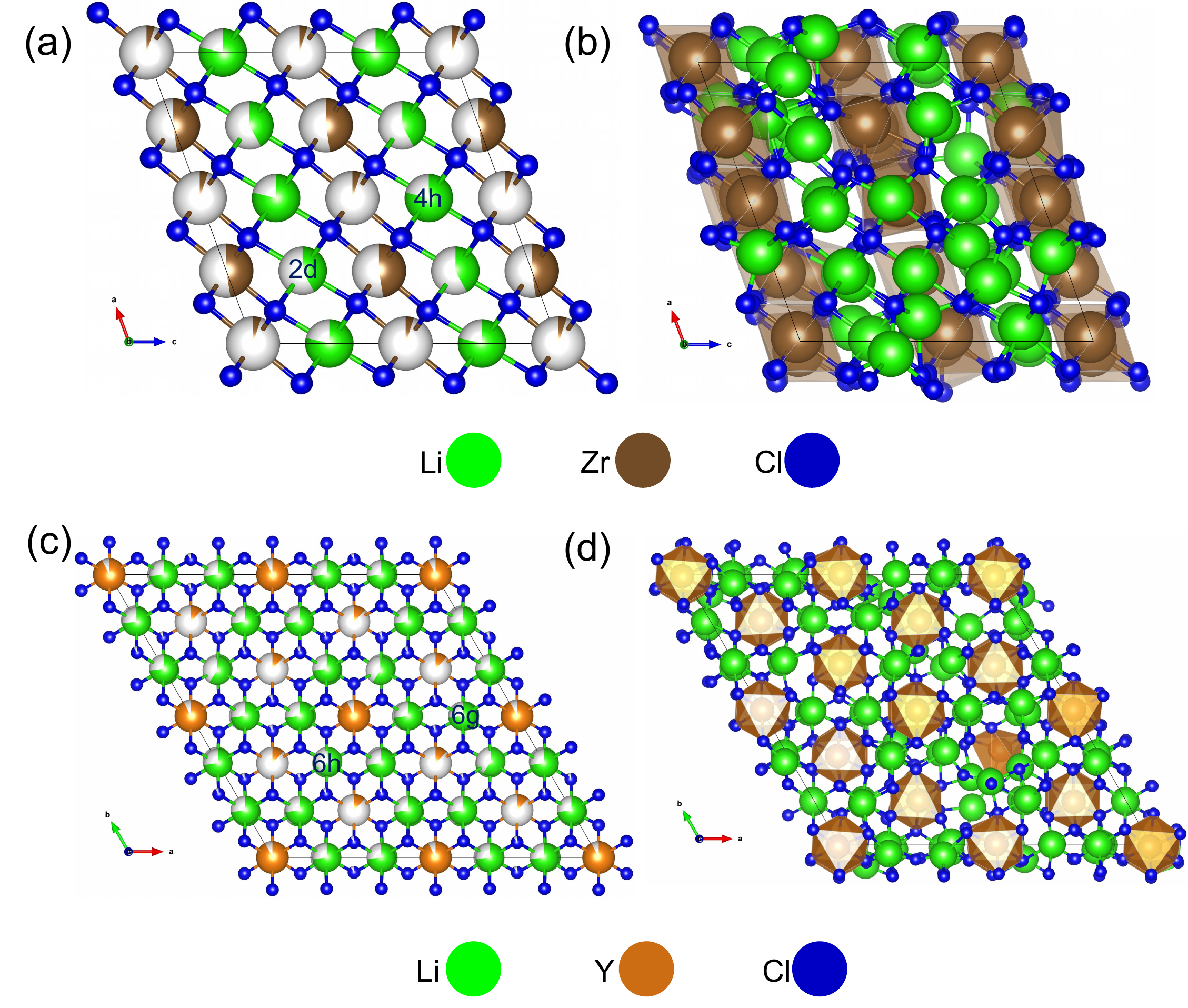}
    \caption{
Crystal structures and Li sublattice environments in halide solid electrolytes.
Panels (a) and (c) show the partially occupied structural models of
Li$_2$ZrCl$_6$ (LZC) and Li$_3$YCl$_6$ (LYC), respectively, following
the reported average structures in Refs.\cite{kwakNewCostEffectiveHalide2021,asanoSolidHalideElectrolytes2018}.
Panels (b) and (d) present the corresponding lowest-energy configurations among
ten $2\times 2 \times 2$ supercells generated for LZC and LYC, respectively, from
the partial-occupancy models in (a) and (c).
In LZC (a,b), Li (green), Zr (brown), and Cl (blue) form a network containing 2d and 4h
Wyckoff Li sites.
In LYC (c,d), Li (green), Y (orange), and Cl (blue) arrange into a framework with 6g and
6h Li positions around YCl$_6$ octahedra.
Polyhedral representations highlight the local coordination environments relevant to
Li-ion transport.}
    \label{structure}
\end{figure}

\autoref{structure}a and c show the crystal structures and Li sublattices of the two halide solid electrolytes studied in this work, LZC and LYC, respectively. We modeled the heat-treated structures of both electrolytes because well-defined refined structures are available for them.\cite{asanoSolidHalideElectrolytes2018,kwakNewCostEffectiveHalide2021} The LZC used in the mixing experiments\cite{wang1+122025} and in our Raman measurements was instead prepared by ball milling, which yields a trigonal, hcp-based phase with higher conductivity than monoclinic LZC.\cite{kwakNewCostEffectiveHalide2021} Because both polymorphs consist of octahedrally coordinated \ce{ZrCl6} units,\cite{kwakNewCostEffectiveHalide2021} we expect the local defect chemistry and Cl dynamics identified here to be qualitatively transferable between them.

To obtain well-defined pristine structures for defect calculations, we constructed $2\times2\times2$ supercells of LZC and LYC in which the cation octahedral sites are filled according to their partial occupancies. We generated ten distinct configurations for each electrolyte by randomly assigning atoms to the partially occupied Wyckoff sites according to their refined occupancies, rounding the number of atoms on each site type to the nearest
integer and leaving the remaining sites vacant.
This procedure yielded 144-atom (32 Li) supercells for LZC and 232-atom (68 Li) supercells for LYC, whose subdomains capture configurational disorder. All structures were fully relaxed using DFT, and the lowest-energy configurations are shown in \autoref{structure}b and \ref{structure}d. These ordered supercell models provide realistic starting structures that capture both the common close-packed anion sublattice and the distinct Li connectivity of LZC and LYC, which underlie their different defect energetics and transport behavior.

\subsection{Defects in LZC, LYC, and the space charge layer at the LZC|LYC interface}
We calculated the formation energies of the following native point defects in LZC and LYC: Li vacancies, Li interstitials, Cl vacancies, and
Cl interstitials, each in the $+1$, $-1$, and neutral charge states.
To assess the effects of Li disorder, we examined multiple
symmetry-inequivalent positions for each defect type (e.g., near and far from the
nearest MCl$_6$ octahedron) to capture the spread in defect energetics
arising from local structural variations.

\begin{figure}[htbp]
    \centering
    \includegraphics[width=0.7\textwidth]{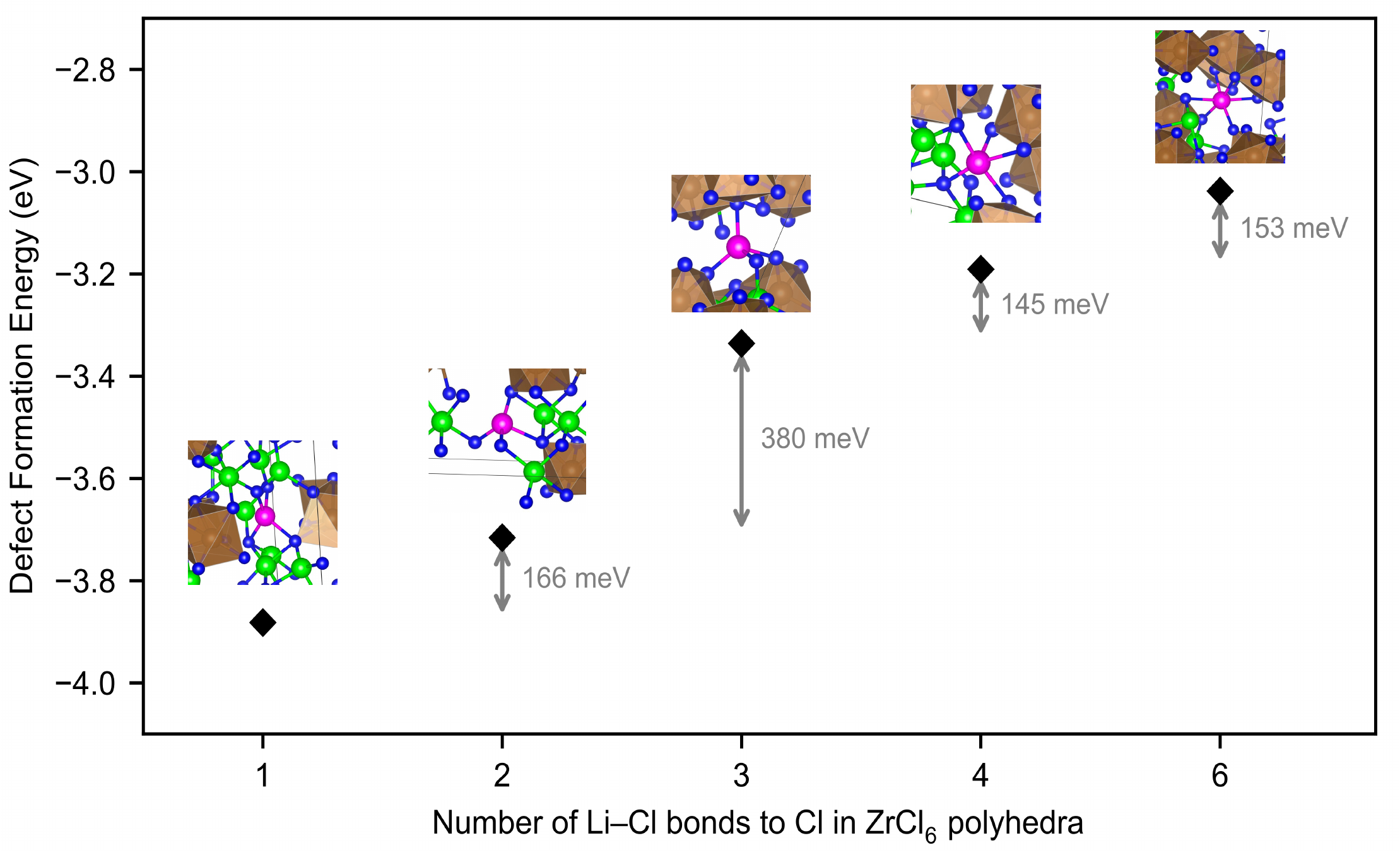}
    \caption{
Defect formation energy (DFE) for \ce{Li+} interstitial configurations in Li$_2$ZrCl$_6$.
The DFE is plotted as a function of the number of bonds the Li interstitial forms with Cl atoms belonging to the surrounding ZrCl$_6$ octahedra. Structural insets
illustrate representative local environments for each coordination motif. Interstitial
configurations in which \ce{Li+} is bonded to \emph{fewer} Cl atoms from ZrCl$_6$
octahedra exhibit lower formation energies, whereas bonding to more such Cl atoms
raises the formation energy, revealing a clear structure--energy relationship associated with
octahedral connectivity.
}
    \label{ladder}
\end{figure}

To clarify how the local environment controls the stability of Li interstitials, we computed the formation energies of \ce{Li+} interstitials in five distinct coordination configurations within the LZC supercell. In \autoref{ladder}, these formation energies are plotted against the number of bonds between the interstitial Li and Cl atoms belonging to ZrCl$_6$ octahedra, using a Li--Cl bond-length cutoff of 2.92~\AA. The defect formation energy increases monotonically with the number of such Li--Cl bonds.
This trend demonstrates that tighter embedding of \ce{Li+} interstitials into the ZrCl$_6$ network is
energetically costly and that interstitials prefer more weakly coordinated sites, farther from Zr, in the LZC framework. Because the distance from the transition metal strongly affects defect energetics, we distinguish defects located near and far from the transition metal M (M = Zr, Y) and denote them with the subscripts $nM$ and $fM$, respectively.

\begin{figure}[htbp]
    \centering
    \includegraphics[width=0.7\textwidth]{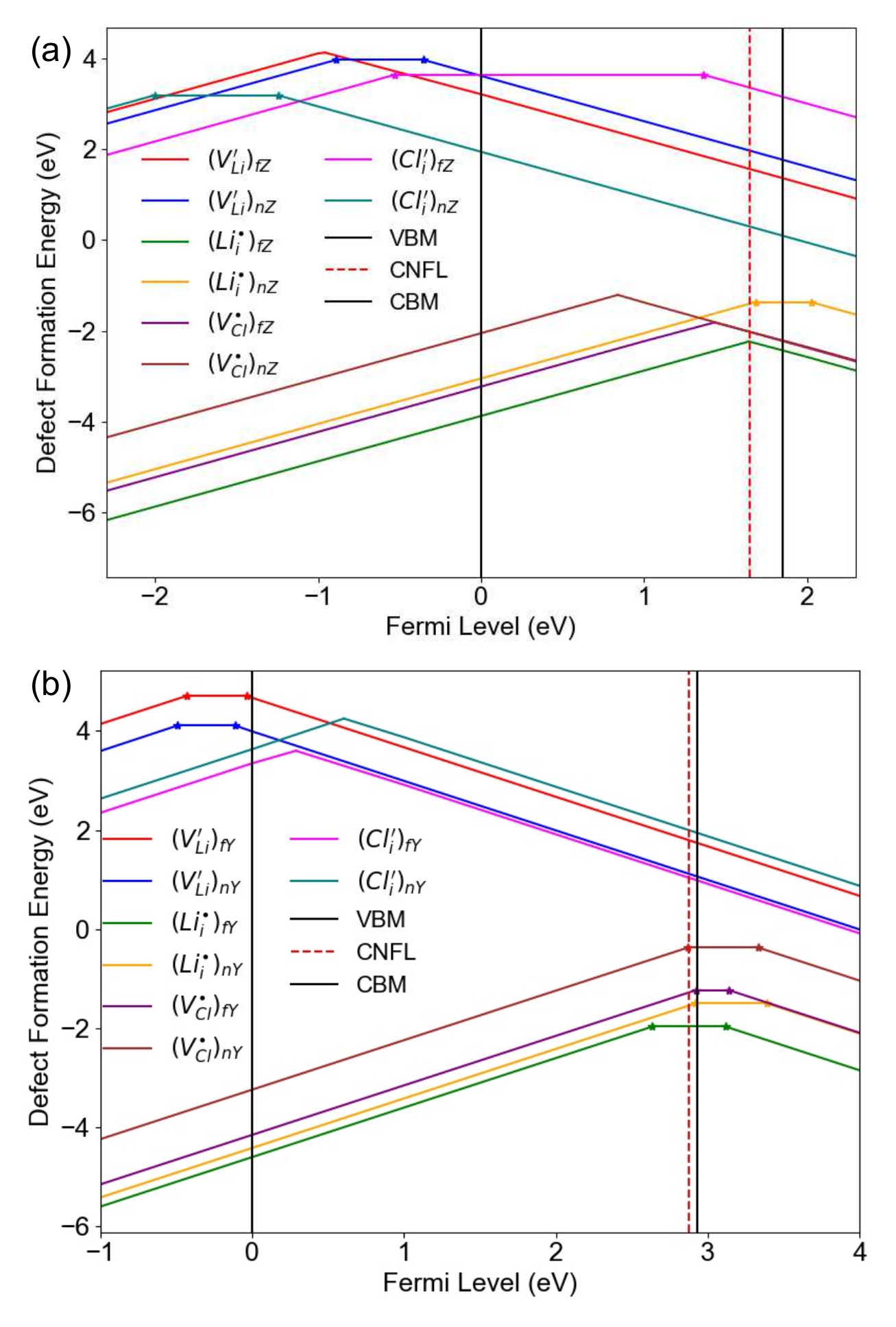}
\vspace{-17pt}
    \caption{
Formation energies of native point defects in (a) \ce{Li2ZrCl6} (LZC) and (b) \ce{Li3YCl6} (LYC) as a function of the Fermi level. Defects include Li vacancies ($V_{\mathrm{Li}}^{'}$), Li interstitials ($\mathrm{Li}_{i}^{\bullet}$), Cl vacancies ($V_{\mathrm{Cl}}^{\bullet}$), and Cl interstitials ($\mathrm{Cl}_{i}^{'}$) in both near- and far-neighbor configurations. In LZC and LYC, the subscripts $nZ$ and $fZ$, and $nY$ and $fY$, respectively, denote defect sites located closer to or farther from the nearest Zr and Y atoms, as indicated in the legends. Solid black vertical lines mark the valence-band maximum (VBM) and the conduction-band minimum (CBM), while the dashed red vertical line marks the charge-neutrality Fermi level (CNFL).
}

    \label{DFE-fermi}
\end{figure}

\autoref{DFE-fermi} shows the formation energies of Li and Cl interstitials and vacancies in LZC and LYC as a function of the Fermi level. In LZC, the most stable donor and acceptor defects are $(Li_i^{\bullet})_{fZ}$ and $(Cl_i^{\prime})_{nZ}$, respectively, whereas in LYC they are $(Li_i^{\bullet})_{fY}$ and $(Cl_i^{\prime})_{fY}$, respectively.
These low DFEs imply that \ce{Li+} interstitials can be readily generated and that \ce{Cl-} interstitials can act as effective charge-compensating defects, together providing a dense population of mobile Li carriers. Consequently, \ce{Li+} and \ce{Cl-} interstitials are expected to play central roles both in mediating fast ion transport and in shaping the space charge profile across the LZC|LYC heterointerface. Considering only Li defects may lead to incorrect predictions of the space charge layers at the interface, as has been pointed out for the LiPON/\ce{Li_xV2O5} interface.\cite{pustorinoMultiionDefectModel2026}

\begin{sidewaystable}[htbp]
\centering
\caption{Possible defect exchange reactions at the interface between LZC and LYC,
constructed using the lowest-energy near/far configurations of each charged
defect at the respective charge-neutrality Fermi levels (CNFLs).}
\label{tab:defect_exchange}

\small
\setlength{\tabcolsep}{1pt}
\renewcommand{\arraystretch}{1.25}

\begin{tabular}{
c
>{\centering\arraybackslash}p{10.0cm}
>{\centering\arraybackslash}p{10.0cm}
c
}
\hline
\textbf{Name} &
\textbf{Reaction} &
\textbf{$\Delta E$ (formula)} &
\textbf{$\Delta E$ (eV)} \\
\hline

R1 &
$\mathrm{Li_{Li}^{\times}[LZC]}
\rightarrow
(V_{\mathrm{Li}}^{\prime})_{fZ}[\mathrm{LZC}]
+
(Li_i^{\bullet})_{fY}[\mathrm{LYC}]$
&
$E^f[(V_{\mathrm{Li}}^{\prime})_{fZ}^{\mathrm{LZC}}]
+
E^f[(Li_i^{\bullet})_{fY}^{\mathrm{LYC}}]$
&
$-0.16$ \\

R2 &
$\mathrm{Li_{Li}^{\times}[LYC]}
\rightarrow
(V_{\mathrm{Li}}^{\prime})_{nY}[\mathrm{LYC}]
+
(Li_i^{\bullet})_{fZ}[\mathrm{LZC}]$
&
$E^f[(V_{\mathrm{Li}}^{\prime})_{nY}^{\mathrm{LYC}}]
+
E^f[(Li_i^{\bullet})_{fZ}^{\mathrm{LZC}}]$
&
$-1.12$ \\

R3 &
$(V_{\mathrm{Li}}^{\prime})_{fZ}[\mathrm{LZC}]
+
\mathrm{Li_{Li}^{\times}[LYC]}
\rightarrow
\mathrm{Li_{Li}^{\times}[LZC]}
+
(V_{\mathrm{Li}}^{\prime})_{nY}[\mathrm{LYC}]$
&
$-E^f[(V_{\mathrm{Li}}^{\prime})_{fZ}^{\mathrm{LZC}}]
+
E^f[(V_{\mathrm{Li}}^{\prime})_{nY}^{\mathrm{LYC}}]$
&
$-0.46$ \\

R4 &
$(V_{\mathrm{Li}}^{\prime})_{nY}[\mathrm{LYC}]
+
\mathrm{Li_{Li}^{\times}[LZC]}
\rightarrow
\mathrm{Li_{Li}^{\times}[LYC]}
+
(V_{\mathrm{Li}}^{\prime})_{fZ}[\mathrm{LZC}]$
&
$-E^f[(V_{\mathrm{Li}}^{\prime})_{nY}^{\mathrm{LYC}}]
+
E^f[(V_{\mathrm{Li}}^{\prime})_{fZ}^{\mathrm{LZC}}]$
&
$0.46$ \\

R5 &
$(Li_i^{\bullet})_{fZ}[\mathrm{LZC}]
\rightarrow
(Li_i^{\bullet})_{fY}[\mathrm{LYC}]$
&
$-E^f[(Li_i^{\bullet})_{fZ}^{\mathrm{LZC}}]
+
E^f[(Li_i^{\bullet})_{fY}^{\mathrm{LYC}}]$
&
$0.50$ \\

R6 &
$(Li_i^{\bullet})_{fY}[\mathrm{LYC}]
\rightarrow
(Li_i^{\bullet})_{fZ}[\mathrm{LZC}]$
&
$-E^f[(Li_i^{\bullet})_{fY}^{\mathrm{LYC}}]
+
E^f[(Li_i^{\bullet})_{fZ}^{\mathrm{LZC}}]$
&
$-0.50$ \\

\hline

R7 &
$\mathrm{Cl_{Cl}^{\times}[LZC]}
\rightarrow
(V_{\mathrm{Cl}}^{\bullet})_{nZ}[\mathrm{LZC}]
+
(Cl_i^{\prime})_{fY}[\mathrm{LYC}]$
&
$E^f[(V_{\mathrm{Cl}}^{\bullet})_{nZ}^{\mathrm{LZC}}]
+
E^f[(Cl_i^{\prime})_{fY}^{\mathrm{LYC}}]$
&
$-0.99$ \\

R8 &
$\mathrm{Cl_{Cl}^{\times}[LYC]}
\rightarrow
(V_{\mathrm{Cl}}^{\bullet})_{fY}[\mathrm{LYC}]
+
(Cl_i^{\prime})_{nZ}[\mathrm{LZC}]$
&
$E^f[(V_{\mathrm{Cl}}^{\bullet})_{fY}^{\mathrm{LYC}}]
+
E^f[(Cl_i^{\prime})_{nZ}^{\mathrm{LZC}}]$
&
$-0.98$ \\

R9 &
$(V_{\mathrm{Cl}}^{\bullet})_{nZ}[\mathrm{LZC}]
+
\mathrm{Cl_{Cl}^{\times}[LYC]}
\rightarrow
\mathrm{Cl_{Cl}^{\times}[LZC]}
+
(V_{\mathrm{Cl}}^{\bullet})_{fY}[\mathrm{LYC}]$
&
$-E^f[(V_{\mathrm{Cl}}^{\bullet})_{nZ}^{\mathrm{LZC}}]
+
E^f[(V_{\mathrm{Cl}}^{\bullet})_{fY}^{\mathrm{LYC}}]$
&
$0.75$ \\

R10 &
$(V_{\mathrm{Cl}}^{\bullet})_{fY}[\mathrm{LYC}]
+
\mathrm{Cl_{Cl}^{\times}[LZC]}
\rightarrow
\mathrm{Cl_{Cl}^{\times}[LYC]}
+
(V_{\mathrm{Cl}}^{\bullet})_{nZ}[\mathrm{LZC}]$
&
$-E^f[(V_{\mathrm{Cl}}^{\bullet})_{fY}^{\mathrm{LYC}}]
+
E^f[(V_{\mathrm{Cl}}^{\bullet})_{nZ}^{\mathrm{LZC}}]$
&
$-0.75$ \\

R11 &
$(Cl_i^{\prime})_{nZ}[\mathrm{LZC}]
\rightarrow
(Cl_i^{\prime})_{fY}[\mathrm{LYC}]$
&
$-E^f[(Cl_i^{\prime})_{nZ}^{\mathrm{LZC}}]
+
E^f[(Cl_i^{\prime})_{fY}^{\mathrm{LYC}}]$
&
$0.74$ \\

R12 &
$(Cl_i^{\prime})_{fY}[\mathrm{LYC}]
\rightarrow
(Cl_i^{\prime})_{nZ}[\mathrm{LZC}]$
&
$-E^f[(Cl_i^{\prime})_{fY}^{\mathrm{LYC}}]
+
E^f[(Cl_i^{\prime})_{nZ}^{\mathrm{LZC}}]$
&
$-0.74$ \\

\hline
\end{tabular}

\end{sidewaystable}

Using the calculated defect formation energies, we next determine how defects redistribute when an interface forms between LZC and LYC. Defect exchange reactions at the interface give rise to space charge layers, which can enhance or impede Li-ion transport depending on the concentration and mobility of the Li defects.\cite{maierIonicConductionSpace1995,wang1+122025,ahmadUnifiedConsistentElectrical2025,limonHeterogeneityPointDefect2024c} Following previous studies,\cite{zhangSynergeticEffectsInorganic2016,panDesignNanostructuredHeterogeneous2016} we use bulk defect formation energies in this analysis; determining explicit interfacial effects on defect formation and exchange is computationally expensive and will be the subject of future work.\cite{ahmadInterfacesSolidElectrolyte2021,limonHeterogeneityPointDefect2024c}
We consider two types of reactions: (i) formation of interfacial Frenkel pairs, in which an ion moves from a lattice site in one phase to an interstitial site in the other, and (ii) transfer of a vacancy or an interstitial from one phase to the other.
\autoref{tab:defect_exchange} lists the possible reactions involving the dominant defects, together with their reaction energies $\Delta E$ evaluated at the CNFL. A more exhaustive analysis covering interfacial Frenkel pair formation for all Li and Cl defect sites (\textit{nM} and \textit{fM}) is given in Tables S1 and S2.
Among the Li-related reactions, R1, R2, R3, and R6 are thermodynamically favorable ($\Delta E < 0$). R2 has the strongest driving force ($\Delta E = -1.12$~eV): Li moves from a lattice site in LYC to an interstitial site in LZC, creating a Li vacancy in LYC and a Li interstitial in LZC. R2 therefore increases the concentration of mobile charge carriers on both sides of the interface.
The transfer reactions point in the same direction: R6 transfers a Li interstitial from LYC to LZC ($\Delta E = -0.50$~eV), and R3 transfers a Li vacancy from LZC to LYC ($\Delta E = -0.46$~eV). Together, R2, R3, and R6 indicate a consistent thermodynamic driving force for Li to redistribute from LYC toward LZC.
Among the Cl-related reactions, R7, R8, R10, and R12 are thermodynamically favorable. The negative reaction energies of R7 and R8 ($\Delta E = -0.99$ and $-0.98$~eV, respectively) indicate that interfacial Cl Frenkel pairs form readily starting from Cl in either phase. R10 ($\Delta E = -0.75$~eV) and R12 ($\Delta E = -0.74$~eV) correspond to favorable transfer of Cl vacancies and Cl interstitials, respectively, from LYC to LZC. Both reactions increase the Cl defect concentrations in LZC.

Collectively, the defect exchange reactions predict a space charge layer consisting of a Li-interstitial-enriched region on the LZC side and a Li-vacancy-enriched region on the LYC side of the interface. The resulting increase in Li concentration on the LZC side is expected to raise the conductivity near the interface. This interpretation is supported by previous studies of \ce{Fe^3+}- and \ce{Ga^3+}-doped LZC, in which aliovalent substitution of \ce{Zr^4+} increases the Li concentration and, in turn, the Li-ion conductivity.\cite{kwakNewCostEffectiveHalide2021,wuEngineeringLi2ZrCl6Ga32026}

\subsection{Simulations of Li-ion transport and dynamics}
\begin{figure}[htbp]
    \centering
    \includegraphics[width=\textwidth]{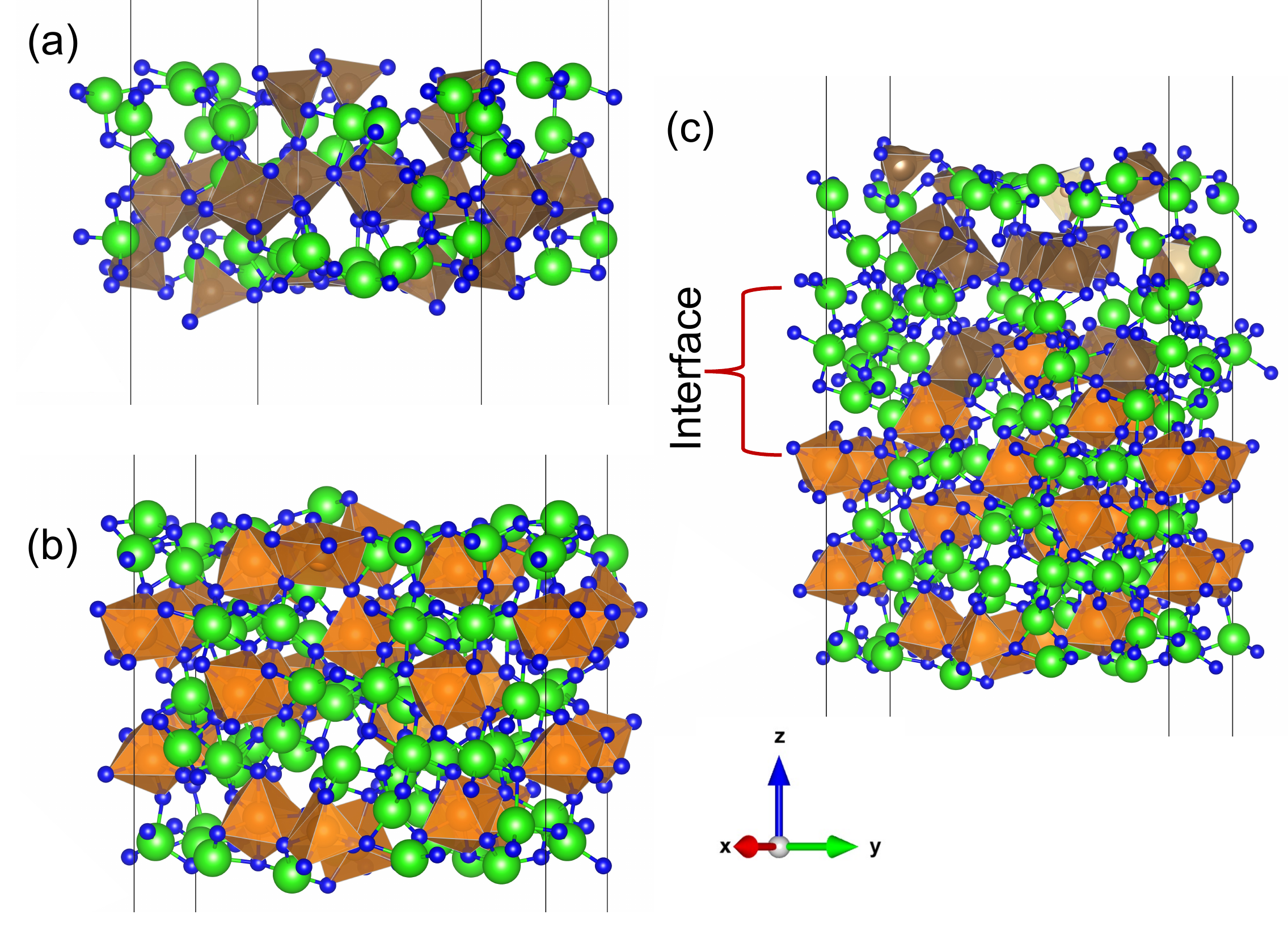}
    \caption{
Atomistic models of the LZC slab, the LYC slab, and the LZC$\vert$LYC heterointerface.
(a) Relaxed LZC slab generated from the lowest-energy bulk LZC configuration.
(b) Relaxed LYC slab generated from the lowest-energy bulk LYC configuration.
(c) Fully relaxed LZC$\vert$LYC interface supercell constructed by joining the LZC and LYC slabs
shown in (a) and (b). The $x$ and $y$ axes of the interface correspond to the $a$ and $b$ axes of LZC and the $c$ and $a$ axes of LYC, chosen to minimize strain (see SI Note 1).
The interfacial region, indicated by the bracket, exhibits an
interdigitated network of adjacent ZrCl$_6$ (brown) and YCl$_6$ (orange) polyhedra, which maintain connectivity across the boundary with modest local tilts and
distortions.
}

    \label{slab_intrfc}
\end{figure}

In addition to defect exchange, the crystal structure and octahedral distortion at the interface can also modulate the Li-ion conductivity.
To probe these structural factors, we constructed atomistic models of the LZC|LYC interface. We first built surface slabs from the lowest-energy bulk configurations of LZC and LYC. The in-plane lattice parameters ($a$ and $b$ for LZC; $c$ and $a$ for LYC) were then matched by applying a small biaxial strain to each phase ($<3\%$ for both), and the atomic positions of the resulting slabs were fully relaxed (\autoref{slab_intrfc}a and b). In both slabs, the bulk-like networks of ZrCl$_6$ and YCl$_6$ octahedra are largely preserved, with only slight distortion near the surface layers. The slabs were then joined to form an LZC|LYC interface, which was further relaxed to give the final interface model shown in \autoref{slab_intrfc}c. Hereafter, $z$ denotes the axis perpendicular to the interface, and $x$ and $y$ denote the in-plane axes.
At the interface, the ZrCl$_6$ (brown) and YCl$_6$ (orange) polyhedra form an interdigitated,
corner-sharing network across the boundary rather than a sharply segregated
stacking of the two phases. The octahedra undergo only modest tilts and local
distortions confined to a few layers around the interface, indicating that no
significant structural reconstruction or amorphization is required to accommodate
the small lattice mismatch. The computed adhesion energy of $-0.47~\mathrm{J\,m^{-2}}$ shows that forming the LZC|LYC
interface is energetically favorable relative to the separated slabs,
supporting this configuration as a realistic
structural platform for examining defect chemistry and Li-ion transport
across the halide--halide heterointerface.

To access the longer time and length scales needed to simulate Li transport, we trained an equivariant MLIP based on NequIP\cite{batznerE3EquivariantGraphNeural2022} and used it to perform MD simulations of the bulk phases and the interface. The training dataset consisted of 2,062 DFT configurations sampled from the relaxed LZC slab, LYC slab, and LZC$\vert$LYC interface structures described above (see Methods for details).
The MLIP was deployed in LAMMPS\cite{thompson2022lammps} to perform finite-temperature MD simulations in the NVT ensemble.
From these trajectories, we computed the mean-squared displacement (MSD) of Li atoms,
extracted Li self-diffusion coefficients, and obtained
Li-ion conductivities from the Nernst--Einstein equation.

In addition to the isolated LZC and LYC slabs and the single LZC$\vert$LYC interface,
we constructed a periodic LYC$\vert$LZC superlattice to sample a broader range of
interfacial environments. This superlattice was built by sandwiching a $2\times 2 \times 6$ supercell of LZC between two $2 \times 2 \times 2$ supercells of LYC along the interface normal, resulting in an 896-atom model that contains two equivalent LYC$\vert$LZC interfaces separated by bulk-like regions of each phase (Fig. S5). The local interfacial structures in this superlattice
are consistent with those in the slab-based LZC$\vert$LYC interface
(\autoref{slab_intrfc}c), but the extended periodic geometry provides more
representative sampling of interfacial coordination motifs and Li environments. It also separates the bulk-like regions well from the interfaces, enabling a
more precise characterization of both bulk and interfacial properties. For the MD simulations, we used $4 \times 4 \times 4$ supercells of bulk LZC and LYC (1,152 and 1,856 atoms, respectively) and replicated the superlattice $2 \times 2 \times 1$ times to obtain a 3,584-atom interface model. These larger cells reduce finite-size effects and improve the statistical sampling of Li diffusion, enabling reliable MSD and conductivity calculations. The radial distribution functions (RDFs) of bulk LZC, bulk LYC, and the interface are plotted in Fig. S7. The interface exhibits similar RDF as the bulk halides but shows a broadened Cl--Cl peak, indicating changes in the Cl distribution around the transition metal cations at the interface.

\begin{figure}[htbp]
    \centering
    \includegraphics[width=0.55\textwidth]{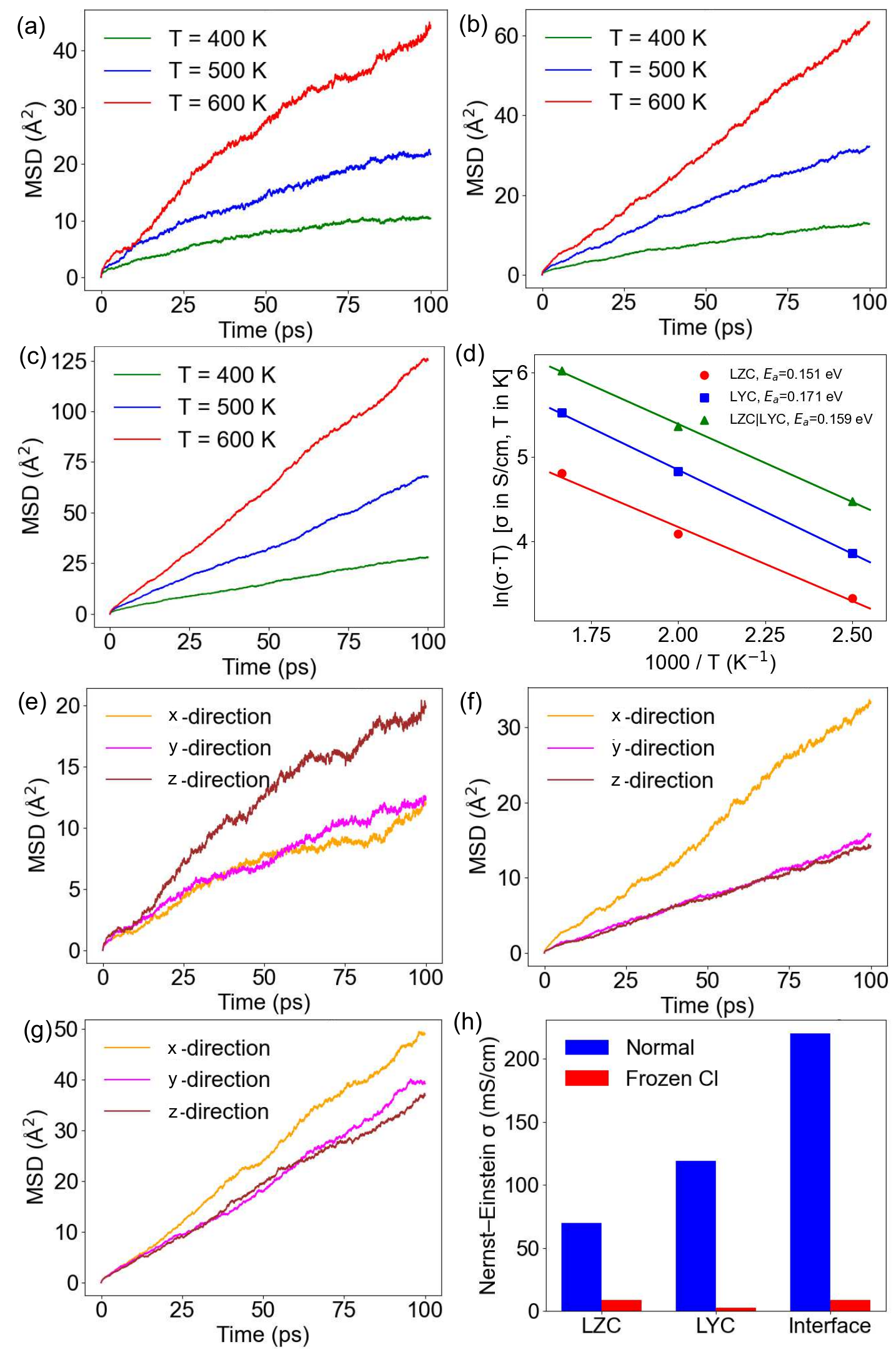}
    \caption{
Li-ion transport in bulk LZC, bulk LYC, and the
LZC$\vert$LYC interface superlattice.
(a--c) Total mean-squared displacement (MSD) of Li versus time at 400, 500, and
600~K for (a) bulk LZC, (b) bulk LYC, and (c) the LZC$\vert$LYC interface. The interface exhibits higher MSD than both LZC and LYC at all temperatures. (d) Arrhenius plots of $\ln(\sigma T)$ as a function of $1000/T$ for the three systems, showing thermally activated behavior. Linear fits yield activation
energies of approximately 0.15~eV for LZC, 0.17~eV for LYC, and 0.16~eV for the
LZC$\vert$LYC interface. (e--g) Direction-resolved MSD of Li at 600~K along the Cartesian
$x$-, $y$-, and $z$-directions defined by the simulation-cell orientation in
\autoref{slab_intrfc} for (e) LZC, (f) LYC, and (g) the
LZC$\vert$LYC interface. The $z$-direction is normal to the interface, while the $x$- and
$y$-directions lie within the interface plane. These Cartesian directions are used consistently across all three systems and differ from the crystallographic axes of bulk LZC ($xy \leftrightarrow ab$) and LYC ($xy \leftrightarrow ca$). (h) Nernst--Einstein Li-ion conductivity at 400~K for LZC, LYC, and the LZC$\vert$LYC interface, comparing simulations with fully mobile ions
(``Normal'') to simulations in which the Cl sublattice is frozen. The drastic reduction in conductivity upon freezing Cl indicates that halide-framework dynamics play a key role in facilitating Li transport, particularly at the interface.
}

    \label{MD_msd}
\end{figure}

\autoref{MD_msd} shows the results of MD simulations of bulk LZC, bulk LYC, and the LZC|LYC interface at 400, 500, and 600~K. The Li MSD increases linearly with time at all temperatures in all three systems (\autoref{MD_msd}a--c), confirming diffusive Li motion. At a given temperature, the Li MSD is smallest in LZC, larger in LYC,
and largest at the interface, indicating higher Li
mobility in the heterostructure. Arrhenius analysis of the Nernst--Einstein conductivities (\autoref{MD_msd}d) yields activation energies of approximately 0.15~eV for
LZC, 0.17~eV for LYC, and 0.16~eV for the interface. Thus,
the enhanced transport at the interface arises primarily from a larger pre-exponential factor (a greater mobile carrier density and/or more available pathways) rather than from a reduced migration barrier. The space charge layer is expected to increase the pre-exponential factor at the interface further by generating additional defects available for transport.
MD simulations of LZC and LYC containing the defects ($\mathrm{Li}_i$, $V_\mathrm{Cl}$, and $V_\mathrm{Li}$) show that these defects can slightly lower the activation energy, thereby enhancing the conductivity (Fig. S6).

The direction-resolved MSDs at 600~K (\autoref{MD_msd}e--g) reveal distinct anisotropies in the three systems. In both LZC and LYC, Li diffusion is fastest along the crystallographic $c$-direction, whereas motion along the $a$- and $b$-directions is more limited. This anisotropy reflects differences in the potential energy surface for Li migration within the $ab$ plane and along the $c$-direction. It is consistent with previous \textit{ab initio} MD simulations of LYC, which attributed the sluggish transport within the $ab$ plane to the need for Li to pass through tetrahedral sites.\cite{ahammedConfigurationalDisorderStrong2024}
Because the $c$-axis maps onto $z$ in LZC and onto $x$ in LYC in the common interface frame, LZC shows the fastest diffusion along $z$, whereas LYC shows the fastest diffusion along $x$. The anisotropy is considerably reduced at the LZC$\vert$LYC interface, although the $x$-direction retains the highest diffusivity, primarily owing to the contribution from LYC. This reduced anisotropy indicates a more three-dimensional Li migration network, which enhances ionic conductivity.

\autoref{MD_msd}h shows the Nernst--Einstein Li-ion conductivities of LZC, LYC, and the interface at 400~K. The interface model exhibits the highest conductivity, followed by LYC and then LZC. Because the Cl sublattice, rather than the transition metal cation, is known to play a critical role in facilitating Li migration in halide solid electrolytes,\cite{ahammedConfigurationalDisorderStrong2024} we also calculated the conductivity with the Cl sublattice artificially frozen. The conductivity of
all three systems drops by an order of magnitude. This strong sensitivity to halide
motion indicates that collective distortions and fluctuations of the Cl
framework play an essential role in opening and modulating Li migration
pathways, particularly in the structurally mixed environment of the
LZC$\vert$LYC heterointerface.

\begin{figure}[htbp]
    \centering
    \includegraphics[width=\textwidth]{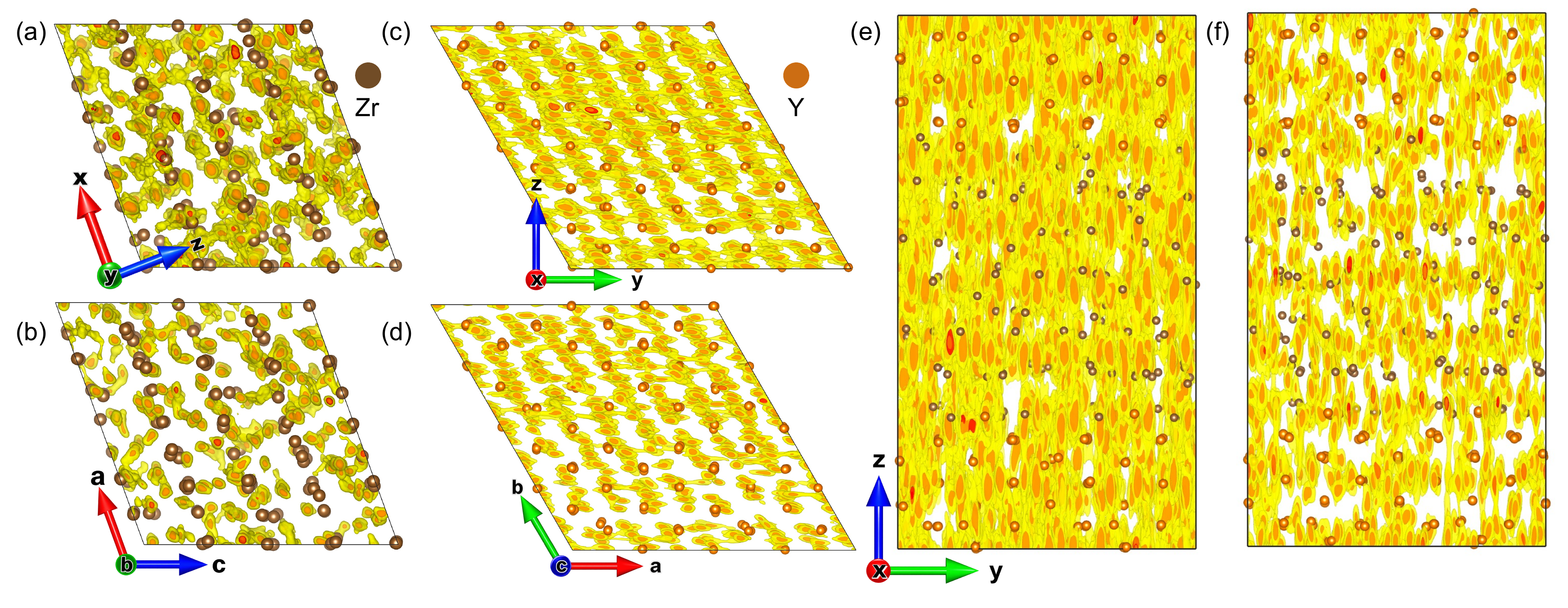}
    \caption{
Li-ion probability-density isosurfaces obtained from MD simulations at
400~K, evaluated over the final 10~ps of the 100~ps production trajectory.
Panels (a) and (b) show LZC with fully mobile and frozen Cl sublattices,
respectively; panels (c) and (d) show the corresponding LYC systems; and
panels (e) and (f) show the LZC$\vert$LYC interface model with fully mobile
and frozen Cl sublattices, respectively. For the bulk LZC and LYC systems,
the crystallographic $a$-, $b$-, and $c$-directions are shown together with
the common Cartesian $x$-, $y$-, and $z$-directions used for directional
transport analysis. For the interface model, the common Cartesian frame is
shown, with $z$ normal to the interface plane and $x$ and $y$ defining the
in-plane directions. The Cartesian directions provide a common reference
frame for comparing Li transport across LZC, LYC, and the interface and do
not necessarily coincide with the crystallographic axes of the individual
bulk phases. Red and yellow isosurfaces denote 50\% and 95\% cumulative Li
probability, respectively. Systems with fully mobile Cl exhibit more extended
and interconnected Li-density pathways, whereas freezing the Cl sublattice
constricts these pathways into more localized regions. This reduction in
accessible Li-transport channels is consistent with the strong suppression
of ionic conductivity observed in \autoref{MD_msd}h, highlighting the
important role of halide-framework dynamics in facilitating Li transport,
particularly at the LZC$\vert$LYC interface.
}

    \label{probability}
\end{figure}

To clarify the microscopic origin of these conductivity trends, we plot Li probability-density isosurfaces in \autoref{probability}. These maps visualize the time-averaged Li occupancy in the bulk phases and at the heterointerface, with red and yellow isosurfaces enclosing 50\% and 95\% cumulative probability, respectively. With fully mobile Cl, both LZC and LYC (\autoref{probability}a and c) display extended, filamentary probability networks, reflecting interconnected migration channels formed by collective distortions of the halide framework. These pathways are especially pronounced in the LZC|LYC interface model (\autoref{probability}e), where the interdigitated ZrCl$_6$/YCl$_6$ network produces a structurally mixed environment that enables Li to percolate laterally within the interface plane while still accessing sites along the interface normal. This broadened Li probability landscape is consistent with the enhanced in-plane diffusivity in \autoref{MD_msd}g and the higher overall conductivity of the interface relative to the bulk
phases.

In contrast, freezing the Cl sublattice dramatically changes the Li probability
landscape (\autoref{probability}b, d, and f). The filamentary networks collapse into isolated pockets in all three systems, indicating that Li motion becomes confined to near-equilibrium wells
with few accessible transition pathways. This strong localization of the
probability density mirrors the order-of-magnitude reduction in
conductivity in \autoref{MD_msd}h and demonstrates that
halide-framework dynamics are essential for opening and sustaining low-barrier
migration channels. The effect is most pronounced at the LZC$\vert$LYC
interface, where suppressing collective Cl motion removes the structural
flexibility required to maintain its mixed percolation pathways.
These results show that the enhanced Li-ion mobility at the
LZC|LYC heterointerface arises from a synergy between the interdigitated
octahedral connectivity and the dynamic fluctuations of the halide framework,
which together generate a dense manifold of migration pathways that is lost when Cl dynamics are constrained.

\subsection{Role of halide ion dynamics}

\begin{figure}[htbp]
    \centering
    \includegraphics[width=0.7\textwidth]{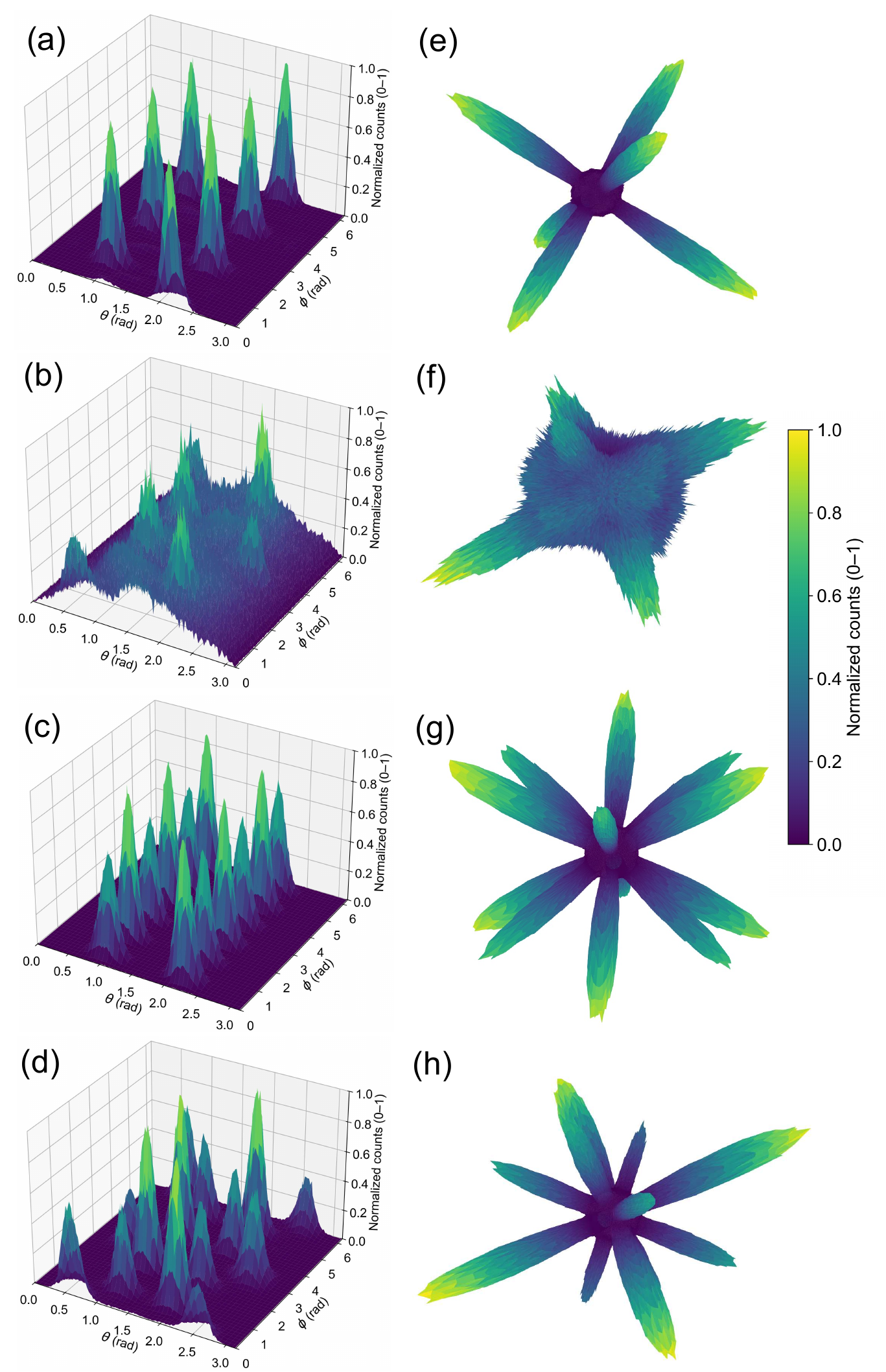}
    \caption{
    Directional distributions of Cl around the framework cations, presented as angular histograms and corresponding spherical plots. Panels (a--d) show the normalized $(\theta,\phi)$ histograms of Cl positions within a cutoff sphere around all framework cations (Y or Zr) for (a) bulk LYC, (b) bulk LZC, (c) LYC at the LYC$\vert$LZC interface, and (d) LZC at the LYC$\vert$LZC interface. Panels (e--h) show the same distributions as spherical plots, with color indicating the normalized occupation (0--1) on a common scale: (e) corresponds to (a), (f) to (b), (g) to (c), and (h) to (d).}

    \label{solid_angle}
\end{figure}

To further understand how Cl dynamics facilitate Li transport in the bulk electrolytes and at the interface, we performed a statistical analysis of Cl positions from the MD simulations. We collected all Cl position vectors relative to the framework cations (Y or Zr) within a cutoff of 3.0~\AA{} over the 100~ps production run and plotted the resulting distributions of polar ($\theta$) and azimuthal
($\phi$) angles as histograms and spherical plots in \autoref{solid_angle}.
The histograms in \autoref{solid_angle}a--d are normalized to their respective maxima to enable direct comparison of anisotropy patterns, while the 3D renderings in \autoref{solid_angle}e--h map the same normalized intensity onto a sphere, providing an intuitive view of the preferred coordination directions.

The bulk phases exhibit highly structured distributions with multiple sharp lobes, indicating well-defined, symmetry-imposed directions of Cl coordination around the framework cation. LYC (\autoref{solid_angle}a and e) displays six pronounced maxima at specific polar and azimuthal angles, whereas LZC (\autoref{solid_angle}b and f) exhibits broader peaks, indicating a wider spread of accessible Cl orientations around Zr. These differences show that the two halide frameworks impose distinct angular constraints on the local anion arrangement.

\autoref{solid_angle}c and d show the angular distributions of Cl in the LYC and LZC layers, respectively, of the interface model. The distributions remain anisotropic, with localized high-probability peaks, but new lobes appear that correspond to additional stable Cl positions. Because more stable positions are accessible near the interface, Cl needs to rearrange less to accommodate a Li hop. These results suggest a shallower potential energy landscape for Cl displacement at the interface, which is expected to facilitate Li-ion hopping through the coupling between framework motion and Li migration.

\subsection{Mechanism of Li-ion migration}

\begin{figure}[htbp]
    \centering
    \includegraphics[width=\textwidth]{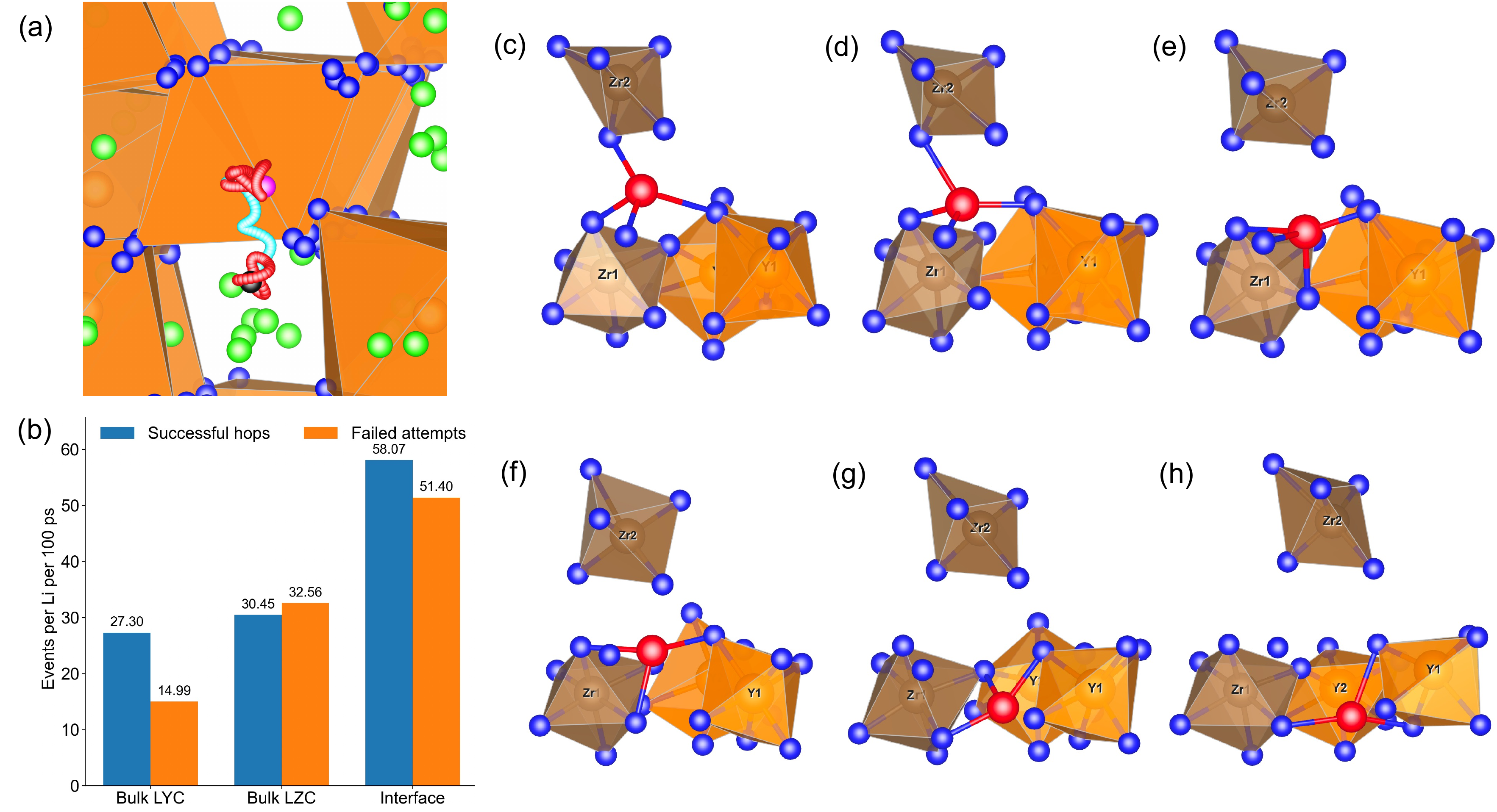}
    \caption{
Li-ion hop classification and atomistic migration mechanism at the LZC|LYC interface.
(a) Representative trajectory of a single Li-ion comparing an unsuccessful attempt (red) with a successful hop (cyan). Unsuccessful attempts are shown by placing the Li-ion at both the initial and attempted final positions in red. For reference, the initial position is marked in black and the final position in magenta; other Li-ions are shown in green and Cl ions in blue.
(b) Event rates (per Li per 100 ps) for successful hops (blue) and unsuccessful attempts (orange) in bulk LYC, bulk LZC, and at the LZC$\vert$LYC interface, highlighting the enhanced hopping activity at the interface.
(c--h) Step-by-step hop mechanism for a Li-ion (red) migrating from a Zr-adjacent site in LZC toward the LZC$\vert$LYC interface. During migration, the Li-ion undergoes a sequence of coordination changes, breaking Li--Cl bonds in its initial environment and forming new ones as it moves between neighboring sites, illustrating how local anion coordination facilitates interfacial transport.
}

    \label{hop}
\end{figure}

We next analyze the MD trajectories to determine the mechanism of Li migration near the interface. \autoref{hop} shows Li positions during representative successful and unsuccessful hopping events. \autoref{hop}a illustrates the distinction between the two: in an unsuccessful attempt, the ion explores a neighboring site but relaxes back (red), whereas a true site-to-site transition occurs only when the ion crosses the bottleneck and stabilizes in the adjacent coordination environment (cyan). A large fraction of the trajectory involves short-range motion around a stable site, and only a subset of these excursions develop into full hops.

Consistent with this microscopic picture, the hop and attempt rates in \autoref{hop}b differ clearly between the bulk phases and the interface. The interface exhibits many more attempts and hops than the bulk phases. The fraction of successful attempts for the interface lies between those of LZC and LYC, as the simulated interface model includes contributions from hopping within bulk layers as well. This observation agrees with the earlier result showing negligible change in activation energy for the interface model. The more flexible Cl environment at the interface likely promotes more attempts.

\autoref{hop}c--h show the step-by-step mechanism of a representative Li hop that originates at a Zr-adjacent site in LZC and progresses toward the LZC|LYC interface.
As the Li-ion leaves its initial site, it undergoes a sequence of coordination rearrangements in which Li--Cl bonds are progressively weakened and broken, followed by the formation of new Li--Cl bonds that stabilize the intermediate configurations and the final site.
This bond-exchange process proceeds through transient under- and re-coordinated states and reflects the ability of the anion sublattice to reorganize locally to accommodate Li translation between neighboring sites.
Importantly, as the ion approaches the interface, the mixed local environment and heterogeneous polyhedral network provide additional accessible configurations for Li--Cl rebonding. This interpretation is further supported by the new stable orientations of Cl relative to the framework cations that appear at the interface but not in the bulk (\autoref{solid_angle}g and h).

\subsection{Vibrational softening at the LZC|LYC interface}

\begin{figure}[htbp]
    \centering
    \includegraphics[width=0.9\textwidth]{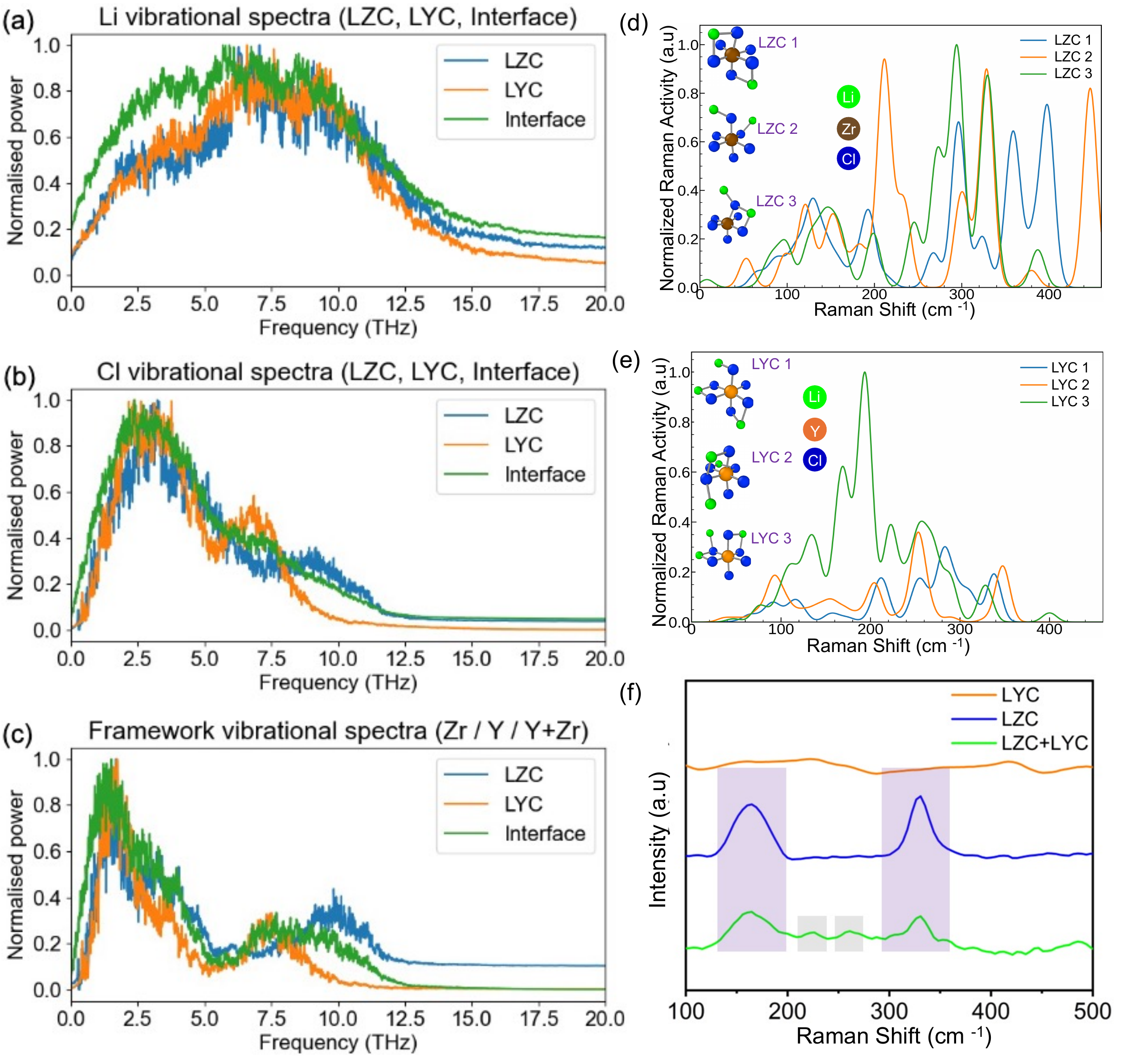}
    \caption{Vibrational power spectra obtained from velocity autocorrelation functions at 400~K for
    (a) Li, (b) Cl, and (c) framework species (Zr for LZC, Y for LYC, and Zr$+$Y combined for the
    LZC$\vert$LYC superlattice), computed from three separate MD
    trajectories: bulk LZC, bulk LYC, and the LZC$\vert$LYC interface model.
    All spectra are normalized for comparison. Relative to both bulk phases, the interface exhibits enhanced low-frequency ($<4$~THz) Li spectral weight, accompanied by softening and broadening of the Cl and framework spectra and increased overlap of Li, Cl, and framework modes in the 1--4~THz range, consistent with the enhanced interfacial Li diffusivity in the transport simulations. (d,e) Simulated Raman spectra of three randomly selected local structures from the 400~K production runs of (d) bulk LZC and (e) bulk LYC. Each spectrum was calculated from a representative local coordination cluster and normalized to compare relative peak positions and spectral features. (f) Experimental Raman spectra of LYC, LZC, and a 1:1 molar mixture of LZC and LYC (LZC$+$LYC), showing the characteristic Raman signatures of the bulk phases and their mixture.
    }

    \label{power_spec}
\end{figure}

In superionic conductors, fast ion transport is often coupled to the vibrational dynamics of the host framework; the vibrational spectrum therefore offers a direct window into the mechanisms that enable ion migration. For LYC, \citet{ahammedConfigurationalDisorderStrong2024} showed that Li diffusion is enabled by the simultaneous activation of low-frequency, Cl-dominated framework modes (1--3~THz) and higher-frequency, Li-dominated modes that move Li into the diffusion channels. The low-frequency modes involve coupled motion of Li, Y, and Cl, are strongly anharmonic with lifetimes of about 0.1~ps or shorter, produce large-amplitude Li oscillations, and overlap strongly with the saddle-point configurations of Li hops. Strikingly, freezing the Cl sublattice in their simulations removed all vibrations below about 3~THz and nearly fully suppressed Li diffusion, demonstrating that framework vibrations are essential for Li transport in LYC. On this basis, they proposed that lattice softness, acting together with configurational disorder, can serve as a design principle for superionic halide conductors.\cite{ahammedConfigurationalDisorderStrong2024}
These findings suggest that changes in the low-frequency vibrational spectrum should accompany, and may help explain, the enhanced Li mobility at the LZC|LYC interface. The local rearrangement of the Cl environment near the interface (\autoref{solid_angle}) is expected to modify the force constants that govern these coupled Li--framework modes. Because strong anharmonicity limits the validity of harmonic phonon calculations in these materials,\cite{ahammedConfigurationalDisorderStrong2024} we computed species-resolved vibrational spectra directly from the finite-temperature MD trajectories, which naturally include anharmonic effects.

\autoref{power_spec}a--c compare the species-resolved (Li, Cl, and transition metal cation) power spectra of bulk LZC, bulk LYC, and the interface, obtained from velocity autocorrelation functions of the MD trajectories. The spectra reveal pronounced differences in lattice dynamics between the bulk phases and the interface. For Li, Cl, and the framework cations
(Zr and Y), the interface exhibits increased
low-frequency spectral weight compared with bulk LZC and LYC.
This redistribution is most evident in the Li spectrum, where the
interface shows enhanced intensity below $\sim$4~THz, consistent with
softening of the local vibrational environment for Li motion at the
heterointerface. Because similar low-frequency coupled vibrations of
Li and the Cl framework facilitate ionic hopping in LYC,\cite{ahammedConfigurationalDisorderStrong2024} this low-frequency enhancement points to a softer
interfacial vibrational environment that is favorable for Li-ion transport.

The Cl-projected spectrum further supports this interpretation. In bulk LZC and LYC, high-frequency Cl vibrations appear near $\sim$9~THz and $\sim$7~THz, respectively. At the interface, these high-frequency contributions are reduced, while the low-frequency Cl spectral weight below $\sim$2.5~THz increases. This indicates softening and broadening of the Cl vibrations upon interface formation, consistent with local distortion of the native Zr--Cl and Y--Cl coordination environments. A similar low-frequency redistribution appears in the framework-projected spectrum, indicating increased overlap among Li, Cl, and framework vibrational modes in the low-frequency regime. These changes provide a lattice-dynamical basis for the enhanced Li mobility observed in the diffusivity analysis.

To probe the vibrations of the \ce{ZrCl6} and \ce{YCl6} octahedra, we also simulated the Raman spectra of bulk LZC and LYC and measured the Raman spectra of LZC, LYC, and their mixture (\autoref{power_spec}d--f). \autoref{power_spec}d and e show Raman spectra simulated for local environments around the framework cation (shown in the insets), extracted from three randomly selected snapshots of the 400~K production MD runs of bulk LZC and LYC, respectively. \autoref{power_spec}f shows the measured Raman spectra of LZC, LYC, and an equimolar LZC--LYC mixture. Because the mixture contains LZC|LYC interfaces, its Raman spectrum is expected to contain vibrational signatures of both bulk phases and their interfaces.

The simulated Raman spectra of LZC show dominant features in the $\sim$100--200~cm$^{-1}$ and $\sim$300--350~cm$^{-1}$ regions, consistent with the measured spectrum of LZC in \autoref{power_spec}f. The characteristic stretching and bending modes of $[\mathrm{ZrCl}_6]^{2-}$ at $\sim$330 and $\sim$161~cm$^{-1}$, respectively, are present in both pristine LZC and the LZC$+$LYC mixture, indicating that the $[\mathrm{ZrCl}_6]^{2-}$ units remain intact at the interface.\cite{kwakNewCostEffectiveHalide2021,beltran2026mechanochemistry} For LYC, the simulated spectra show stronger features primarily in the $\sim$200--300~cm$^{-1}$ region, whereas these features are very weak in the measured spectrum. For the LZC--LYC mixture, we observe peak broadening that can be attributed to local distortion of the $[(\mathrm{ZrCl}_{4/2})\mathrm{Cl}_2]_n$ structural chains.\cite{am1998vibrational} The intensity near $\sim$226 and $\sim$261~cm$^{-1}$ in the mixture can be assigned to $[\mathrm{YCl}_6]^{3-}$ vibrational modes, particularly at the interface.\cite{brahmbhatt2023synthesis,papatheodorou1977raman} Together, the broadening of the LZC peaks and the appearance of new modes at intermediate wavenumbers indicate a modification of the Cl sublattice rather than a simple coexistence of $[\mathrm{ZrCl}_6]^{2-}$ and $[\mathrm{YCl}_6]^{3-}$ octahedra. These findings are consistent with the formation of a distorted network of bridged $\mathrm{ZrCl}_6/\mathrm{YCl}_6$ octahedra at the LZC$\vert$LYC interface upon mixing.

\section{Conclusions}

We have established a mechanistic understanding of Li-ion transport in halide solid electrolytes and determined how it is modified at their interfaces by two key factors: space charge layer formation and distortion of the \ce{MCl6} octahedra. The complementary techniques employed in this work reveal several phenomena that collectively determine the Li-ion conductivity of bulk LZC, bulk LYC, and the LZC--LYC mixture, including defect exchange at the interface, changes in Li site connectivity and migration statistics, the rearrangement and orientation of Cl around the framework cations, and changes in the vibrational spectra. First-principles defect calculations reveal that defect formation energies are highly sensitive to the distance from the transition metal cation (\autoref{ladder}). Defect thermodynamics further reveals the
creation of new charge carriers at the interface, enriching LYC with Li vacancies and LZC with Li interstitials. Our MD simulations highlight the critical role of Cl rearrangement in facilitating Li migration, showing that collective halide motion is essential for sustaining percolating Li transport channels. The appearance of new angular positions for Cl around the framework cations clearly indicates that Cl rearranges more easily at the interface. The LZC|LYC heterointerface also promotes in-plane Li diffusion along the mixed \ce{ZrCl6}/\ce{YCl6} network and reduces transport anisotropy, providing three-dimensional pathways for Li conduction. Finally, the vibrational spectra show a shift of the Li and Cl modes to lower frequencies at the interface, indicating lattice softening that is favorable for Li motion.

Our work shows that multicomponent systems provide a promising platform for enhancing the conductivity of solid electrolytes through their heterointerfaces. A difference in the Li chemical potentials of two electrolytes can drive Li transfer across their interface, generating defects (e.g., through Frenkel pair formation) that support faster transport. Although both the space charge layer and octahedral distortion contribute to the conductivity enhancement, their relative contributions remain to be quantified. It would be valuable to examine how these contributions change with different metal dopants, halide anions, or processing routes.
Furthermore, how space charge layer formation affects octahedral distortion, and vice versa, is an important open question with implications for the maximum conductivity achievable in multicomponent systems. Future investigations along these lines may enable substantial improvements in the conductivity of solid electrolytes, helping them simultaneously satisfy the stringent requirements of solid-state batteries.

\section{Methods}

\subsection*{Density Functional Theory (DFT) Calculations}
We investigated Li$_2$ZrCl$_6$ (LZC), Li$_3$YCl$_6$ (LYC), and the coherent LZC$|$LYC heterointerface using density functional theory (DFT) as implemented in Quantum Espresso.\cite{giannozziQUANTUMESPRESSOModular2009,giannozziAdvancedCapabilitiesMaterials2017} Total energies and forces were computed using the Perdew--Burke--Ernzerhof (PBE) exchange--correlation functional\cite{perdewGeneralizedGradientApproximation1996} with ultrasoft pseudopotentials.\cite{garrityPseudopotentialsHighthroughputDFT2014} Brillouin-zone integrations employed a $2\times2\times2$ \textit{k}-point grid. Plane-wave kinetic energy cutoffs were set to \texttt{ecutwfc} = 50~Ry and \texttt{ecutrho} = 250~Ry. Structural relaxations were performed until total-energy changes were below
\texttt{etot\_conv\_thr} = $1.0\times10^{-4}$~Ry and residual forces were below
\texttt{forc\_conv\_thr} = $1.0\times10^{-3}$~Ry/Bohr. The DFT-D3 dispersion correction of Grimme and coworkers\cite{Grimme2010} was used to account for van der Waals interactions. Supercells, surfaces, and interfaces were generated using pymatgen\cite{Ong2013pymatgen} and the Atomic Simulation Environment (ASE),\cite{HjorthLarsen2017atomic} and structures were visualized using VESTA.\cite{Momma2011vesta} The total density of states for LZC and LYC are shown in Fig. S2.

Initial LZC and LYC structures were constructed from reported X-ray diffraction (XRD) refinements.\cite{kwakNewCostEffectiveHalide2021,asanoSolidHalideElectrolytes2018} Based on the refined site occupancies, we generated ten independent $2\times2\times2$ supercells for each phase by randomly assigning atoms to the partially occupied sites, followed by full structural relaxation. The lowest-energy relaxed configuration of each phase was used as the pristine reference structure for subsequent defect and interface calculations.

\subsection*{Charged Defect Formation Energies}
We computed defect formation energies (DFEs) for Li vacancies, Li interstitials, Cl vacancies, and Cl interstitials in the $+1$, $-1$, and neutral charge states. The DFE of a defect in charge state $q$ was evaluated as\cite{stegmaierLiDefectsSolidState2017,freysoldtFullyInitioFiniteSize2009,kim2020quick,limonHeterogeneityPointDefect2024c}
\begin{equation}
E_f(\epsilon_F,\{\mu_i\}) =
E_d - E_p - \sum_i n_i\mu_i + q(\epsilon_{\mathrm{VBM}}+\epsilon_F) + \Delta_{\mathrm{corr}},
\label{dfe_corr}
\end{equation}
where $E_d$ and $E_p$ are the total energies of the defective and pristine supercells, respectively; $n_i$ is the number of atoms of species $i$ (Li or Cl) added ($n_i=+1$ for an interstitial) or removed ($n_i=-1$ for a vacancy); $\mu_i$ is the chemical potential of species $i$; $q$ is the defect charge state; $\epsilon_{\mathrm{VBM}}$ is the energy of the valence-band maximum (VBM); and $\epsilon_F$ is the Fermi level referenced to the VBM.

The term $\Delta_{\mathrm{corr}}$ corrects for spurious electrostatic contributions arising from the compensating background charge and from interactions between periodic images of the defect in charged bulk supercells. These corrections were computed following the scheme of Freysoldt and coworkers\cite{freysoldtFullyInitioFiniteSize2009,freysoldt2018first} using \texttt{sxdefectalign}.\cite{freysoldtFullyInitioFiniteSize2009}

The required dielectric tensors were computed using density-functional perturbation theory as implemented in the \texttt{ph.x} and \texttt{dynmat.x} modules of Quantum Espresso.\cite{giannozziQUANTUMESPRESSOModular2009,giannozziAdvancedCapabilitiesMaterials2017} The dielectric tensors of LZC and LYC are provided in Fig. S1.

\subsubsection*{Defect Site Sampling}
For LZC, Li vacancies were created on both crystallographic Li sites (2$d$ and 4$h$) to identify the energetically preferred vacancy site, and five distinct initial Li interstitial placements were tested and fully relaxed. For LYC, both Li sites (6$g$ and 6$h$) were considered for vacancy formation and candidate interstitial placements. The lowest-energy relaxed configuration of each defect type was retained for analysis.

\subsection*{Interface Model Construction and Energetics}
Coherent LZC and LYC slabs were constructed from the lowest-energy bulk configurations, and lattice matching was achieved by selecting commensurate in-plane supercells that minimize strain. The heterointerface was assembled by joining the relaxed slabs along the interface normal, after which the atomic positions were fully relaxed with the in-plane lattice vectors held fixed at the matched values. The interface formation (adhesion) energy was computed as
\begin{equation}
E_{\mathrm{interface}}=\frac{1}{A}\left(E_{\mathrm{total}}-n_1E_{\mathrm{LZC}}-n_2E_{\mathrm{LYC}} \right),
\end{equation}
where $E_{\mathrm{total}}$ is the total energy of the interface supercell, $E_{\mathrm{LZC}}$ and $E_{\mathrm{LYC}}$ are the energies of the corresponding pristine reference slabs (constructed with the same in-plane lattice constants as the interface), $n_1$ and $n_2$ are the numbers of formula units in the LZC and LYC portions of the interface cell, and $A$ is the interfacial area. The relaxed interface structure and the full lattice-matching and strain metrics are summarized in the SI in Supplementary Note 1 and Fig. S5.

\subsection*{Machine-Learned Interatomic Potential (NequIP)}
To access longer time and length scales for Li transport at near-DFT accuracy, we
constructed an MLIP using the NequIP equivariant
neural-network framework.\cite{batznerE3EquivariantGraphNeural2022} The training dataset comprised
2\,062 DFT-labeled configurations sampled from the relaxed LZC slab, LYC slab, and
LZC$\vert$LYC interface structures described above, including snapshots with finite-temperature atomic displacements to sample the local energy and force landscape relevant to molecular dynamics. Only slab and interface geometries
were included because all corresponding DFT calculations employed a dipole correction
consistent with the interfacial geometry; bulk configurations computed without this
correction were excluded to maintain a uniform reference. The dataset was randomly
split into 80\,\% training, 10\,\% validation, and 10\,\% test sets, and the model was
trained on total energies and atomic forces until convergence of the validation loss.

\subsection*{Molecular Dynamics Simulations}

All large-scale MD simulations were performed in the NVT ensemble in LAMMPS\cite{thompson2022lammps} using the
trained NequIP potential. In addition to the $2\times2\times2$ supercell-based LZC and LYC slab models and the corresponding slab-based LZC$\vert$LYC interface, we constructed a periodic LYC$\vert$LZC superlattice to sample a broader range of interfacial environments.
The superlattice was generated by stacking three $2\times2\times2$ LZC supercells
between two $2\times2\times2$ LYC supercells along the interface normal, yielding an
896-atom periodic model containing two equivalent LYC$\vert$LZC interfaces separated by
bulk-like regions of each phase (Fig. S5). The local
interfacial coordination motifs in this superlattice are consistent with those in the
slab-based interface model (\autoref{slab_intrfc}c), while the extended periodic
geometry improves sampling of interfacial environments and enables clear separation of
bulk-like and interfacial regions.

For MD simulations, the $2\times2\times2$ cells used for DFT and NequIP training were
expanded to reduce finite-size effects and improve statistical sampling. The LZC and
LYC $2\times2\times2$ supercells were replicated to $4\times4\times4$ models, and the
LYC$\vert$LZC superlattice was replicated $2\times2\times1$ in $(x,y,z)$, resulting in
1\,152-atom LZC, 1\,856-atom LYC, and 3\,584-atom LZC$\vert$LYC superlattice systems.
These larger models preserve the local environments represented in the DFT-trained
NequIP potential while providing sufficient lateral extent for reliable MSD-based
diffusivity and conductivity calculations.

Before the production runs at 400, 500, and 600~K, each system was equilibrated using a
stepwise heating protocol in the NVT ensemble with a time step of 1~fs. Starting from the
minimized structure at 1~K, the temperature was increased in 50~K increments, with
alternating 2~ps holds and 2~ps linear ramps, until reaching the desired target temperature.
This procedure ensured gradual heating and near-equilibrium configurations at all
intermediate steps.

After reaching 400, 500, or 600~K, each system was equilibrated at the target temperature
and then simulated for 100~ps in the NVT ensemble. Unwrapped atomic coordinates were used for transport
analysis, and Li self-diffusion coefficients and ionic conductivities were
obtained from the Li MSDs using the Einstein and Nernst--Einstein
relations, respectively, as described below. Details of the ``frozen-Cl''
simulations are given in the Frozen-Anion Simulations subsection.

\subsection*{Diffusivity and Ionic Conductivity Calculations}
Lithium diffusivities were extracted from the long-time slope of the Li mean-squared
displacement (MSD) via the Einstein relation,
\begin{equation}
D_{\mathrm{Li}}=\lim_{t\rightarrow\infty}\frac{1}{2d}\frac{\mathrm{d}}{\mathrm{d}t}\left\langle
|\mathbf{r}_i(t)-\mathbf{r}_i(0)|^2\right\rangle,
\end{equation}
where $d$ is the dimensionality ($d=3$ for isotropic bulk; for directional diffusivities,
$d=1$ with $\mathbf{r}$ projected onto $x$, $y$, or $z$). Ionic conductivity was computed
using the Nernst--Einstein relation,
\begin{equation}
\sigma_{\mathrm{NE}}=\frac{n_{\mathrm{Li}} q^2 D_{\mathrm{Li}}}{k_BT},
\end{equation}
where $n_{\mathrm{Li}}$ is the number density of mobile Li-ions, $q$ is the Li charge,
$k_B$ is the Boltzmann constant, and $T$ is the temperature. No Haven-ratio correction
was applied. Activation energies and prefactors were obtained by fitting the Nernst--Einstein conductivities to the Arrhenius form
$\sigma T=A\exp(-E_a/k_BT)$, using data at
400, 500, and 600~K. Diffusivities and conductivities were computed using multiple
time origins along the trajectory to improve statistical sampling.

\subsection*{Frozen-Anion Simulations}
To isolate the role of halide framework dynamics in Li transport, we performed
``frozen-Cl'' simulations in which the Cl sublattice was immobilized during the
production run. Specifically, Cl atoms were constrained by setting their velocities
and forces to zero, while the remaining atoms (Li and framework cations, Zr/Y) evolved
dynamically. Frozen-Cl simulations were
performed at 400~K and were initialized from equilibrated configurations obtained from
fully mobile (normal) simulations at the same temperature. The Cl sublattice was frozen
only during the production stage, and Li diffusivities and Nernst--Einstein
conductivities were computed using the same MSD-based workflow as for the fully mobile
trajectories.

\subsection*{Li Probability-Density Isosurfaces}
Probability-density maps for Li were computed from NequIP-based NVT trajectories at
400~K. For each system, the final 10~ps of the 100~ps production run (10\,001 frames)
were used to construct a three-dimensional occupation histogram on a uniform 0.1~\AA\
voxel grid. The raw densities were smoothed with a Gaussian kernel of width
$\sigma = 1.5$ voxels, after which each map was normalized such that
$\int \rho(\mathbf{r})\, dV = 1$. Isosurface values defining the 50\% (core) and 95\%
(halo) cumulative probability thresholds were determined separately for each
trajectory (LZC bulk, LYC bulk, and the interface), ensuring that the
visualized regions enclose the same cumulative probability in each system and reflect
intrinsic differences in Li spatial distributions rather than differences in absolute
density scale. All probability analysis was performed in the simulation cell without
drift correction, and the same grid resolution and smoothing parameters were applied to
all systems for consistent comparison.

\subsection*{Vibrational Power Spectra from Velocity Autocorrelation}
Vibrational spectra were computed from NequIP-based NVT trajectories at 400~K using
velocity autocorrelation functions (VACFs) evaluated separately for Li, Cl, and
framework species (Zr for bulk LZC, Y for bulk LYC, and Zr$+$Y for the interface
superlattice). Unwrapped coordinates were used, and for each species the center-of-mass motion was removed on a frame-by-frame basis prior to velocity estimation; additionally, the time-averaged mean of each velocity component was subtracted before computing the VACF. Velocities
were obtained by a central-difference scheme from the species positions, and the
normalized VACF was defined as
\begin{equation}
C_v(t)=\frac{\left\langle \mathbf{v}(t)\cdot\mathbf{v}(0)\right\rangle}
{\left\langle \mathbf{v}(0)\cdot\mathbf{v}(0)\right\rangle},
\end{equation}
where the average is taken over atoms of the given species and Cartesian components.
VACFs were evaluated using the Wiener--Khinchin theorem by Fourier transforming the
velocity time series, multiplying by its complex conjugate, and inverse transforming to
obtain the autocorrelation, followed by normalization to $C_v(0)$.

A species-resolved spectral function was obtained as the real part of the discrete
Fourier transform of $C_v(t)$, with frequencies determined from the sampling interval.
For visualization, each spectrum was shifted to be nonnegative and normalized by its
maximum value to enable direct comparison of spectral shapes across systems.

\subsection*{Simulated Raman Spectra}
Representative local coordination clusters centered on the framework cation
(Zr for LZC and Y for LYC) were extracted from three randomly selected
structures along the 400~K production MD trajectories. Raman calculations were
performed using ORCA~\cite{ORCA,ORCA5} with the PBE0 hybrid density functional~\cite{Adamo1999}, Grimme's D3BJ
dispersion correction~\cite{Grimme2010,Grimme2011}, and the def2-TZVP basis set~\cite{Weigend2005}. Numerical vibrational
frequencies were evaluated using the \texttt{NUMFREQ} keyword with
\texttt{TightSCF} convergence criteria. Polarizabilities required for Raman
activities were computed using the \texttt{\%ELPROP POLAR 1} option. The
resulting Raman spectra were normalized to their respective maximum intensities
to facilitate comparison of relative peak positions and spectral features among
the sampled local environments.

\subsection*{Local Coordination Anisotropy and Angular Mapping}
To quantify how the local anion environment differs between bulk and interfacial
regions, we analyzed the directional distribution of Cl neighbors around framework
cations (Y in LYC and Zr in LZC) using LAMMPS trajectory dumps. For each trajectory, we
collected displacement vectors from every framework center (all Y or all Zr atoms in
each frame) to all Cl atoms within a cutoff radius of $r_{\mathrm{cut}}=3.0$~\AA. Periodic
boundary conditions were enforced using a minimum-image convention implemented in
fractional coordinates, which is valid for both orthorhombic and triclinic simulation
cells.

Each retained Cl displacement vector $\mathbf{r}=(x,y,z)$ was converted to spherical
angles using $\theta=\arccos\!\left(z/|\mathbf{r}|\right)$ and
$\phi=\mathrm{atan2}(y,x)$ with $\phi$ mapped to $[0,2\pi)$. Orientations collected
from all centers and sampled frames were accumulated into two-dimensional
$(\theta,\phi)$ histograms using 144 bins in $\theta$ over $[0,\pi]$ and 288 bins in
$\phi$ over $[0,2\pi]$. To emphasize anisotropy patterns, each histogram was normalized
by its maximum count to obtain a 0--1 intensity map, and a common 0--1 color scale was
used across all systems.

The normalized $(\theta,\phi)$ distributions were visualized both as solid-angle
histogram surfaces and as 3D ``blob'' renderings on a unit sphere. For the blob
renderings, the radial coordinate was mapped from the normalized intensity $I(\theta,\phi)$
as $R(\theta,\phi)=R_0 + A\,I(\theta,\phi)^{\gamma}$ with $R_0=0.12$, $A=0.88$, and
$\gamma=1.0$, while the surface color encoded the same normalized intensity using a
consistent colormap.

\subsection*{Li Hopping Event Classification}
Li hopping statistics were quantified using a site-based analysis that distinguishes
short-lived cage excursions (unsuccessful attempts) from committed site-to-site
transitions (successful hops). For each trajectory (bulk LYC, bulk LZC, and the
interface superlattice), Li ``sites'' were identified directly from the Li positional
probability density by accumulating Li positions in fractional coordinates on a
$64\times64\times64$ grid, smoothing the resulting histogram with a Gaussian kernel
($\sigma=1$ bin), and detecting local maxima with height at least $0.12$ of the global
maximum. Peaks separated by less than 0.035 in fractional distance (under periodic
wrapping) were merged, yielding a set of $K$ site centers for each system (bulk LYC:
$K=852$; bulk LZC: $K=376$; interface: $K=857$).

Each Li-ion was assigned a site label at every saved frame by finding the nearest site
center under periodic boundary conditions. To ensure robust nearest-neighbor assignment
in triclinic cells, the fractional site centers were converted to Cartesian coordinates
and replicated across the $3\times3\times3$ set of neighboring periodic images; nearest
sites were then determined using a KD-tree search over these replicated site images.
Li positions were wrapped into the primary cell prior to assignment.

Events were identified by tracking each Li-ion as it departed its current residence
site $A$. Upon the first frame in which the site label differed from $A$, the ion was
considered to be in an excursion. Excursions were classified using a commitment
window of $\tau_{\mathrm{commit}}=100$~fs. If the ion returned to its original site $A$
at any time before commitment, the excursion was recorded as an unsuccessful
attempt. Otherwise, the ion was classified as having undergone a successful hop
$A\rightarrow B$ if, over the subsequent $\tau_{\mathrm{commit}}$ window, a site
$B\neq A$ was the most frequent label and occupied at least 80\% of the frames in the
window. Excursions that were still
unresolved at the end of the trajectory were excluded from the statistics.

Event rates were reported as events per Li per 100~ps by normalizing the total
number of successful hops and unsuccessful attempts by the number of Li-ions in the
system and the analyzed trajectory duration (100~ps). Representative hop mechanisms
(Fig.~\ref{hop}c--h) were selected from the event list and visualized by following the
migrating Li-ion and its local Cl coordination environment along the hop pathway.

\subsection*{Synthesis of Halide Electrolytes and Their Mixture}

LZC was synthesized via ball-milling of a stoichiometric ratio of LiCl (Sigma Aldrich, 99.95\% trace metals basis) and \ce{ZrCl4} ($>$99.5\% trace metals basis) in a Cole-Parmer BM-400 Mixer/Mill for 4~cycles of alternating 15~minutes run, 15~minutes rest, and a 40:1 ball-to-powder mass ratio. Similarly, LYC was synthesized through ball-milling of a stoichiometric ratio of LiCl and \ce{YCl3} (anhydrous (99.9\%-Y) (REQ)), including a 5~wt\% excess of \ce{YCl3}, for 6~cycles of 15~minutes run, 15~minutes rest, and a 30:1 ball-to-powder mass ratio. All powders were handled in an Ar-filled glovebox with \ce{O2} and \ce{H2O} levels below 1~ppm. To prepare the mixture of LZC+LYC, LZC and LYC powders were measured out in a 1:1 mol ratio, sealed into a 100~mL polypropylene (PP) mixing container, and mixed at 2000~rpm for 5~minutes in an AR-100 THINKY Conditioning Mixer. The PP containers were transferred back into the glovebox for further use. The resulting mixture and bulk samples were used for Raman spectroscopy.

\section*{Conflicts of interest}

There are no conflicts to declare.

\section*{Data availability}

The data supporting the findings of this study are available within the article, its Supporting Information, and in the repository \url{https://github.com/ahmad-research-group/LZC-LYC-paper}. Additional computational data, including defect configurations, interface models, molecular dynamics input files, trained NequIP model files, and analysis scripts, will be made available in the online repository upon publication.

\begin{acknowledgement}
We acknowledge alternative energy research funding from the Edward E. Whitacre Jr. College of Engineering at Texas Tech University for supporting this research. We acknowledge the Lonestar6 research allocation (DMR24029) at the Texas Advanced Computing Center (TACC) for providing computational resources that have contributed to the research results reported within this paper. This work also used the Texas Tech REPACSS CPU and GPU resources through allocation MAT250134 from the Advanced Cyberinfrastructure Coordination Ecosystem: Services \& Support (ACCESS) program, which is supported by U.S. National Science Foundation grants \#2138259, \#2138286, \#2138307, \#2137603, and \#2138296.
During the preparation of this work, the authors used Gemini and Claude in order to improve the clarity of writing and grammar. After using these tools, the authors reviewed and edited the content as needed and take full responsibility for the content of the publication.

\end{acknowledgement}
\begin{suppinfo}

Details of ordered supercell generation, interface generation, calculated halide electrolyte properties, defect exchange reactions at the interface, radial distribution functions, and effects of defects on ion transport from MD simulations.

\end{suppinfo}

\bibliography{zotero,manual-refs}

\providecommand{\latin}[1]{#1}
\makeatletter
\providecommand{\doi}
  {\begingroup\let\do\@makeother\dospecials
  \catcode`\{=1 \catcode`\}=2 \doi@aux}
\providecommand{\doi@aux}[1]{\endgroup\texttt{#1}}
\makeatother
\providecommand*\mcitethebibliography{\thebibliography}
\csname @ifundefined\endcsname{endmcitethebibliography}
  {\let\endmcitethebibliography\endthebibliography}{}
\begin{mcitethebibliography}{83}
\providecommand*\natexlab[1]{#1}
\providecommand*\mciteSetBstSublistMode[1]{}
\providecommand*\mciteSetBstMaxWidthForm[2]{}
\providecommand*\mciteBstWouldAddEndPuncttrue
  {\def\EndOfBibitem{\unskip.}}
\providecommand*\mciteBstWouldAddEndPunctfalse
  {\let\EndOfBibitem\relax}
\providecommand*\mciteSetBstMidEndSepPunct[3]{}
\providecommand*\mciteSetBstSublistLabelBeginEnd[3]{}
\providecommand*\EndOfBibitem{}
\mciteSetBstSublistMode{f}
\mciteSetBstMaxWidthForm{subitem}{(\alph{mcitesubitemcount})}
\mciteSetBstSublistLabelBeginEnd
  {\mcitemaxwidthsubitemform\space}
  {\relax}
  {\relax}

\bibitem[Manthiram \latin{et~al.}(2017)Manthiram, Yu, and
  Wang]{manthiram2017lithium}
Manthiram,~A.; Yu,~X.; Wang,~S. Lithium battery chemistries enabled by
  solid-state electrolytes. \emph{Nature Reviews Materials} \textbf{2017},
  \emph{2}, 1--16\relax
\mciteBstWouldAddEndPuncttrue
\mciteSetBstMidEndSepPunct{\mcitedefaultmidpunct}
{\mcitedefaultendpunct}{\mcitedefaultseppunct}\relax
\EndOfBibitem
\bibitem[Du \latin{et~al.}(2025)Du, Zhou, and He]{du2025halide}
Du,~B.; Zhou,~H.; He,~P. Halide lithium conductors: from design and synthesis
  to application for all-solid-state batteries. \emph{ACS Applied Energy
  Materials} \textbf{2025}, \emph{8}, 723--745\relax
\mciteBstWouldAddEndPuncttrue
\mciteSetBstMidEndSepPunct{\mcitedefaultmidpunct}
{\mcitedefaultendpunct}{\mcitedefaultseppunct}\relax
\EndOfBibitem
\bibitem[Janek and Zeier(2016)Janek, and Zeier]{janekSolidFutureBattery2016}
Janek,~J.; Zeier,~W.~G. A Solid Future for Battery Development. \emph{Nature
  Energy} \textbf{2016}, \emph{1}, 1--4\relax
\mciteBstWouldAddEndPuncttrue
\mciteSetBstMidEndSepPunct{\mcitedefaultmidpunct}
{\mcitedefaultendpunct}{\mcitedefaultseppunct}\relax
\EndOfBibitem
\bibitem[Wang \latin{et~al.}(2015)Wang, Richards, Ong, Miara, Kim, Mo, and
  Ceder]{wangDesignPrinciplesSolidstate2015}
Wang,~Y.; Richards,~W.~D.; Ong,~S.~P.; Miara,~L.~J.; Kim,~J.~C.; Mo,~Y.;
  Ceder,~G. Design Principles for Solid-State Lithium Superionic Conductors.
  \emph{Nature Materials} \textbf{2015}, \emph{14}, 1026--1031\relax
\mciteBstWouldAddEndPuncttrue
\mciteSetBstMidEndSepPunct{\mcitedefaultmidpunct}
{\mcitedefaultendpunct}{\mcitedefaultseppunct}\relax
\EndOfBibitem
\bibitem[Kim and Siegel(2019)Kim, and
  Siegel]{kimPredictingWettabilityElectrochemical2019}
Kim,~K.; Siegel,~D.~J. Predicting {{Wettability}} and the {{Electrochemical
  Window}} of {{Lithium-Metal}}/{{Solid Electrolyte Interfaces}}. \emph{ACS
  Applied Materials \& Interfaces} \textbf{2019}, \emph{11}, 39940--39950\relax
\mciteBstWouldAddEndPuncttrue
\mciteSetBstMidEndSepPunct{\mcitedefaultmidpunct}
{\mcitedefaultendpunct}{\mcitedefaultseppunct}\relax
\EndOfBibitem
\bibitem[Xiao \latin{et~al.}(2020)Xiao, Wang, Bo, Kim, Miara, and
  Ceder]{xiaoUnderstandingInterfaceStability2020}
Xiao,~Y.; Wang,~Y.; Bo,~S.-H.; Kim,~J.~C.; Miara,~L.~J.; Ceder,~G.
  Understanding Interface Stability in Solid-State Batteries. \emph{Nature
  Reviews Materials} \textbf{2020}, \emph{5}, 105--126\relax
\mciteBstWouldAddEndPuncttrue
\mciteSetBstMidEndSepPunct{\mcitedefaultmidpunct}
{\mcitedefaultendpunct}{\mcitedefaultseppunct}\relax
\EndOfBibitem
\bibitem[Ahmad and Viswanathan(2017)Ahmad, and
  Viswanathan]{ahmadStabilityElectrodepositionSolidSolid2017}
Ahmad,~Z.; Viswanathan,~V. Stability of {{Electrodeposition}} at {{Solid-Solid
  Interfaces}} and {{Implications}} for {{Metal Anodes}}. \emph{Physical Review
  Letters} \textbf{2017}, \emph{119}, 056003\relax
\mciteBstWouldAddEndPuncttrue
\mciteSetBstMidEndSepPunct{\mcitedefaultmidpunct}
{\mcitedefaultendpunct}{\mcitedefaultseppunct}\relax
\EndOfBibitem
\bibitem[Tu \latin{et~al.}(2020)Tu, {Barroso-Luque}, Shi, and
  Ceder]{tuElectrodepositionMechanicalStability2020}
Tu,~Q.; {Barroso-Luque},~L.; Shi,~T.; Ceder,~G. Electrodeposition and
  {{Mechanical Stability}} at {{Lithium-Solid Electrolyte Interface}} during
  {{Plating}} in {{Solid-State Batteries}}. \emph{Cell Reports Physical
  Science} \textbf{2020}, \emph{1}, 100106\relax
\mciteBstWouldAddEndPuncttrue
\mciteSetBstMidEndSepPunct{\mcitedefaultmidpunct}
{\mcitedefaultendpunct}{\mcitedefaultseppunct}\relax
\EndOfBibitem
\bibitem[Albertus \latin{et~al.}(2018)Albertus, Babinec, Litzelman, and
  Newman]{albertusStatusChallengesEnabling2018}
Albertus,~P.; Babinec,~S.; Litzelman,~S.; Newman,~A. Status and Challenges in
  Enabling the Lithium Metal Electrode for High-Energy and Low-Cost
  Rechargeable Batteries. \emph{Nature Energy} \textbf{2018}, \emph{3},
  16--21\relax
\mciteBstWouldAddEndPuncttrue
\mciteSetBstMidEndSepPunct{\mcitedefaultmidpunct}
{\mcitedefaultendpunct}{\mcitedefaultseppunct}\relax
\EndOfBibitem
\bibitem[Kwak \latin{et~al.}(2021)Kwak, Han, Lyoo, Park, Jung, Han, Kwon, Kim,
  Hong, Nam, and Jung]{kwakNewCostEffectiveHalide2021}
Kwak,~H.; Han,~D.; Lyoo,~J.; Park,~J.; Jung,~S.~H.; Han,~Y.; Kwon,~G.; Kim,~H.;
  Hong,~S.-T.; Nam,~K.-W.; Jung,~Y.~S. New {{Cost-Effective Halide Solid
  Electrolytes}} for {{All-Solid-State Batteries}}: {{Mechanochemically
  Prepared Fe3}}+-{{Substituted Li2ZrCl6}}. \emph{Advanced Energy Materials}
  \textbf{2021}, \emph{11}, 2003190\relax
\mciteBstWouldAddEndPuncttrue
\mciteSetBstMidEndSepPunct{\mcitedefaultmidpunct}
{\mcitedefaultendpunct}{\mcitedefaultseppunct}\relax
\EndOfBibitem
\bibitem[Tan \latin{et~al.}(2019)Tan, Banerjee, Deng, Wu, Nguyen, Doux, Wang,
  Cheng, Ong, Meng, and Chen]{tanEnablingThinFlexible2019}
Tan,~D. H.~S.; Banerjee,~A.; Deng,~Z.; Wu,~E.~A.; Nguyen,~H.; Doux,~J.-M.;
  Wang,~X.; Cheng,~J.-h.; Ong,~S.~P.; Meng,~Y.~S.; Chen,~Z. Enabling {{Thin}}
  and {{Flexible Solid-State Composite Electrolytes}} by the {{Scalable
  Solution Process}}. \emph{ACS Applied Energy Materials} \textbf{2019},
  \emph{2}, 6542--6550\relax
\mciteBstWouldAddEndPuncttrue
\mciteSetBstMidEndSepPunct{\mcitedefaultmidpunct}
{\mcitedefaultendpunct}{\mcitedefaultseppunct}\relax
\EndOfBibitem
\bibitem[Lee \latin{et~al.}(2023)Lee, Jang, Lee, Wu, Chen, Holoubek, Yu, Ham,
  Jeon, Kim, Lee, Song, Meng, and
  Chen]{leePhysioElectrochemicallyDurableDryProcessed2023}
Lee,~D.~J.; Jang,~J.; Lee,~J.-P.; Wu,~J.; Chen,~Y.-T.; Holoubek,~J.; Yu,~K.;
  Ham,~S.-Y.; Jeon,~Y.; Kim,~T.-H.; Lee,~J.~B.; Song,~M.-S.; Meng,~Y.~S.;
  Chen,~Z. Physio-{{Electrochemically Durable Dry-Processed Solid-State
  Electrolyte Films}} for {{All-Solid-State Batteries}}. \emph{Advanced
  Functional Materials} \textbf{2023}, \emph{33}, 2301341\relax
\mciteBstWouldAddEndPuncttrue
\mciteSetBstMidEndSepPunct{\mcitedefaultmidpunct}
{\mcitedefaultendpunct}{\mcitedefaultseppunct}\relax
\EndOfBibitem
\bibitem[Li \latin{et~al.}(2020)Li, Liang, Yang, Adair, Wang, Zhao, and
  Sun]{li2020progress}
Li,~X.; Liang,~J.; Yang,~X.; Adair,~K.~R.; Wang,~C.; Zhao,~F.; Sun,~X. Progress
  and perspectives on halide lithium conductors for all-solid-state lithium
  batteries. \emph{Energy \& Environmental Science} \textbf{2020}, \emph{13},
  1429--1461\relax
\mciteBstWouldAddEndPuncttrue
\mciteSetBstMidEndSepPunct{\mcitedefaultmidpunct}
{\mcitedefaultendpunct}{\mcitedefaultseppunct}\relax
\EndOfBibitem
\bibitem[Liang \latin{et~al.}(2021)Liang, Li, Adair, and Sun]{liang2021metal}
Liang,~J.; Li,~X.; Adair,~K.~R.; Sun,~X. Metal halide superionic conductors for
  all-solid-state batteries. \emph{Accounts of Chemical Research}
  \textbf{2021}, \emph{54}, 1023--1033\relax
\mciteBstWouldAddEndPuncttrue
\mciteSetBstMidEndSepPunct{\mcitedefaultmidpunct}
{\mcitedefaultendpunct}{\mcitedefaultseppunct}\relax
\EndOfBibitem
\bibitem[Wang \latin{et~al.}(2022)Wang, Liang, Kim, and Sun]{wang2022prospects}
Wang,~C.; Liang,~J.; Kim,~J.~T.; Sun,~X. Prospects of halide-based
  all-solid-state batteries: From material design to practical application.
  \emph{Science Advances} \textbf{2022}, \emph{8}, eadc9516\relax
\mciteBstWouldAddEndPuncttrue
\mciteSetBstMidEndSepPunct{\mcitedefaultmidpunct}
{\mcitedefaultendpunct}{\mcitedefaultseppunct}\relax
\EndOfBibitem
\bibitem[Nie \latin{et~al.}(2023)Nie, Hu, and Li]{nie2023halide}
Nie,~X.; Hu,~J.; Li,~C. Halide-based solid electrolytes: The history, progress,
  and challenges. \emph{Interdisciplinary Materials} \textbf{2023}, \emph{2},
  365--389\relax
\mciteBstWouldAddEndPuncttrue
\mciteSetBstMidEndSepPunct{\mcitedefaultmidpunct}
{\mcitedefaultendpunct}{\mcitedefaultseppunct}\relax
\EndOfBibitem
\bibitem[Bouguern \latin{et~al.}(2026)Bouguern, Ningappa, Vishweswariah,
  Kumar~MR, Kanno, and Zaghib]{bouguern2026comparative}
Bouguern,~M.~D.; Ningappa,~N.~G.; Vishweswariah,~K.; Kumar~MR,~A.; Kanno,~R.;
  Zaghib,~K. Comparative Advances in Sulfide and Halide Electrolytes for
  Commercialization of All-Solid-State Lithium Batteries. \emph{Advanced
  Materials} \textbf{2026}, \emph{38}, e13255\relax
\mciteBstWouldAddEndPuncttrue
\mciteSetBstMidEndSepPunct{\mcitedefaultmidpunct}
{\mcitedefaultendpunct}{\mcitedefaultseppunct}\relax
\EndOfBibitem
\bibitem[He \latin{et~al.}(2023)He, Zhang, Xin, Xu, Hu, Wu, Yang, and
  Tian]{he2023halogen}
He,~B.; Zhang,~F.; Xin,~Y.; Xu,~C.; Hu,~X.; Wu,~X.; Yang,~Y.; Tian,~H. Halogen
  chemistry of solid electrolytes in all-solid-state batteries. \emph{Nature
  Reviews Chemistry} \textbf{2023}, \emph{7}, 826--842\relax
\mciteBstWouldAddEndPuncttrue
\mciteSetBstMidEndSepPunct{\mcitedefaultmidpunct}
{\mcitedefaultendpunct}{\mcitedefaultseppunct}\relax
\EndOfBibitem
\bibitem[Asano \latin{et~al.}(2018)Asano, Sakai, Ouchi, Sakaida, Miyazaki, and
  Hasegawa]{asanoSolidHalideElectrolytes2018}
Asano,~T.; Sakai,~A.; Ouchi,~S.; Sakaida,~M.; Miyazaki,~A.; Hasegawa,~S. Solid
  {{Halide Electrolytes}} with {{High Lithium-Ion Conductivity}} for
  {{Application}} in 4 {{V Class Bulk-Type All-Solid-State Batteries}}.
  \emph{Advanced Materials} \textbf{2018}, \emph{30}, 1803075\relax
\mciteBstWouldAddEndPuncttrue
\mciteSetBstMidEndSepPunct{\mcitedefaultmidpunct}
{\mcitedefaultendpunct}{\mcitedefaultseppunct}\relax
\EndOfBibitem
\bibitem[Molaiyan \latin{et~al.}(2025)Molaiyan, Jin, Wang, Dos~Reis, Petnikota,
  Lassi, and Paolella]{molaiyan2025chemistry}
Molaiyan,~P.; Jin,~T.; Wang,~S.; Dos~Reis,~G.~S.; Petnikota,~S.; Lassi,~U.;
  Paolella,~A. The chemistry of halide-based solid electrolytes: unlocking
  advances in solid-state Li-ion batteries. \emph{Chemical Communications}
  \textbf{2025}, \emph{61}, 2846--2857\relax
\mciteBstWouldAddEndPuncttrue
\mciteSetBstMidEndSepPunct{\mcitedefaultmidpunct}
{\mcitedefaultendpunct}{\mcitedefaultseppunct}\relax
\EndOfBibitem
\bibitem[Kwak \latin{et~al.}(2022)Kwak, Wang, Park, Liu, Kim, Choi, Mo, and
  Jung]{kwak2022emerging}
Kwak,~H.; Wang,~S.; Park,~J.; Liu,~Y.; Kim,~K.~T.; Choi,~Y.; Mo,~Y.;
  Jung,~Y.~S. Emerging halide superionic conductors for all-solid-state
  batteries: design, synthesis, and practical applications. \emph{ACS Energy
  Letters} \textbf{2022}, \emph{7}, 1776--1805\relax
\mciteBstWouldAddEndPuncttrue
\mciteSetBstMidEndSepPunct{\mcitedefaultmidpunct}
{\mcitedefaultendpunct}{\mcitedefaultseppunct}\relax
\EndOfBibitem
\bibitem[Li and Du(2025)Li, and Du]{li2025building}
Li,~C.; Du,~Y. Building a better all-solid-state lithium-ion battery with
  halide solid-state electrolyte. \emph{ACS nano} \textbf{2025}, \emph{19},
  4121--4155\relax
\mciteBstWouldAddEndPuncttrue
\mciteSetBstMidEndSepPunct{\mcitedefaultmidpunct}
{\mcitedefaultendpunct}{\mcitedefaultseppunct}\relax
\EndOfBibitem
\bibitem[Tuo \latin{et~al.}(2023)Tuo, Sun, and Liu]{tuo2023recent}
Tuo,~K.; Sun,~C.; Liu,~S. Recent progress in and perspectives on emerging
  halide superionic conductors for all-solid-state batteries.
  \emph{Electrochemical Energy Reviews} \textbf{2023}, \emph{6}, 17\relax
\mciteBstWouldAddEndPuncttrue
\mciteSetBstMidEndSepPunct{\mcitedefaultmidpunct}
{\mcitedefaultendpunct}{\mcitedefaultseppunct}\relax
\EndOfBibitem
\bibitem[Kim \latin{et~al.}(2021)Kim, Park, Jung, Chung, Shim, Wood, and
  Yu]{kim2021material}
Kim,~K.; Park,~D.; Jung,~H.-G.; Chung,~K.~Y.; Shim,~J.~H.; Wood,~B.~C.; Yu,~S.
  Material design strategy for halide solid electrolytes Li3MX6 (X= Cl, Br, and
  I) for all-solid-state high-voltage Li-ion batteries. \emph{Chemistry of
  Materials} \textbf{2021}, \emph{33}, 3669--3677\relax
\mciteBstWouldAddEndPuncttrue
\mciteSetBstMidEndSepPunct{\mcitedefaultmidpunct}
{\mcitedefaultendpunct}{\mcitedefaultseppunct}\relax
\EndOfBibitem
\bibitem[Liang \latin{et~al.}(2020)Liang, Li, Wang, Adair, Li, Zhao, Wang, Hu,
  Zhang, Zhao, \latin{et~al.} others]{liang2020site}
Liang,~J.; Li,~X.; Wang,~S.; Adair,~K.~R.; Li,~W.; Zhao,~Y.; Wang,~C.; Hu,~Y.;
  Zhang,~L.; Zhao,~S.; others Site-occupation-tuned superionic Li x ScCl3+ x
  halide solid electrolytes for all-solid-state batteries. \emph{Journal of the
  American Chemical Society} \textbf{2020}, \emph{142}, 7012--7022\relax
\mciteBstWouldAddEndPuncttrue
\mciteSetBstMidEndSepPunct{\mcitedefaultmidpunct}
{\mcitedefaultendpunct}{\mcitedefaultseppunct}\relax
\EndOfBibitem
\bibitem[Wang \latin{et~al.}(2023)Wang, Wang, Liu, Wu, Yu, Duan, Kim, Huang,
  Wang, Mo, \latin{et~al.} others]{wang2023new}
Wang,~C.; Wang,~S.; Liu,~X.; Wu,~Y.; Yu,~R.; Duan,~H.; Kim,~J.~T.; Huang,~H.;
  Wang,~J.; Mo,~Y.; others New insights into aliovalent substituted halide
  solid electrolytes for cobalt-free all-solid state batteries. \emph{Energy \&
  Environmental Science} \textbf{2023}, \emph{16}, 5136--5143\relax
\mciteBstWouldAddEndPuncttrue
\mciteSetBstMidEndSepPunct{\mcitedefaultmidpunct}
{\mcitedefaultendpunct}{\mcitedefaultseppunct}\relax
\EndOfBibitem
\bibitem[Wang \latin{et~al.}(2024)Wang, Zhou, Wang, Guo, Gong, Yao, Wu, Wang,
  Ganapathy, Bai, \latin{et~al.} others]{wang2024designing}
Wang,~Q.; Zhou,~Y.; Wang,~X.; Guo,~H.; Gong,~S.; Yao,~Z.; Wu,~F.; Wang,~J.;
  Ganapathy,~S.; Bai,~X.; others Designing lithium halide solid electrolytes.
  \emph{Nature communications} \textbf{2024}, \emph{15}, 1050\relax
\mciteBstWouldAddEndPuncttrue
\mciteSetBstMidEndSepPunct{\mcitedefaultmidpunct}
{\mcitedefaultendpunct}{\mcitedefaultseppunct}\relax
\EndOfBibitem
\bibitem[Liu \latin{et~al.}(2026)Liu, Zhang, Liu, Cuillier, Cardon, Murphy, Ji,
  and Chen]{liuStructureLi3YCl6Its2026}
Liu,~J.; Zhang,~Y.; Liu,~Z.; Cuillier,~P.; Cardon,~P.; Murphy,~M.; Ji,~H.;
  Chen,~H. Structure of {{Li3YCl6}} and {{Its Effect}} on {{Lithium Ionic
  Conductivity}}. \emph{Journal of the American Chemical Society}
  \textbf{2026}, \relax
\mciteBstWouldAddEndPunctfalse
\mciteSetBstMidEndSepPunct{\mcitedefaultmidpunct}
{}{\mcitedefaultseppunct}\relax
\EndOfBibitem
\bibitem[Lee and Jung(2024)Lee, and Jung]{lee2024synergistic}
Lee,~Y.~C.; Jung,~S.~C. Synergistic effect of lithium ions for fast diffusion
  in Li2ZrCl6 solid electrolytes. \emph{Electrochimica Acta} \textbf{2024},
  \emph{497}, 144632\relax
\mciteBstWouldAddEndPuncttrue
\mciteSetBstMidEndSepPunct{\mcitedefaultmidpunct}
{\mcitedefaultendpunct}{\mcitedefaultseppunct}\relax
\EndOfBibitem
\bibitem[Zhong \latin{et~al.}(2024)Zhong, Gupta, Deng, Jun, and
  Ceder]{zhongEffectCationDisorder2024}
Zhong,~P.; Gupta,~S.; Deng,~B.; Jun,~K.; Ceder,~G. Effect of {{Cation
  Disorder}} on {{Lithium Transport}} in {{Halide Superionic Conductors}}.
  \emph{ACS Energy Letters} \textbf{2024}, \emph{9}, 2775--2781\relax
\mciteBstWouldAddEndPuncttrue
\mciteSetBstMidEndSepPunct{\mcitedefaultmidpunct}
{\mcitedefaultendpunct}{\mcitedefaultseppunct}\relax
\EndOfBibitem
\bibitem[Wu \latin{et~al.}(2026)Wu, Liu, Huang, Zhao, and
  Zhou]{wuEngineeringLi2ZrCl6Ga32026}
Wu,~Y.; Liu,~Y.; Huang,~X.; Zhao,~S.; Zhou,~N. Engineering {{Li2ZrCl6}} with
  {{Ga3}}+/{{Ge4}}+ {{Doping}} for {{High Ionic Conductivity}} and {{Stable
  Interfaces}} in {{Solid-State Batteries}}. \emph{Inorganic Chemistry}
  \textbf{2026}, \emph{65}, 2850--2862\relax
\mciteBstWouldAddEndPuncttrue
\mciteSetBstMidEndSepPunct{\mcitedefaultmidpunct}
{\mcitedefaultendpunct}{\mcitedefaultseppunct}\relax
\EndOfBibitem
\bibitem[Zhou \latin{et~al.}(2020)Zhou, Kwok, Shyamsunder, Zhang, Wu, and
  Nazar]{zhou2020new}
Zhou,~L.; Kwok,~C.~Y.; Shyamsunder,~A.; Zhang,~Q.; Wu,~X.; Nazar,~L.~F. A new
  halospinel superionic conductor for high-voltage all solid state lithium
  batteries. \emph{Energy \& Environmental Science} \textbf{2020}, \emph{13},
  2056--2063\relax
\mciteBstWouldAddEndPuncttrue
\mciteSetBstMidEndSepPunct{\mcitedefaultmidpunct}
{\mcitedefaultendpunct}{\mcitedefaultseppunct}\relax
\EndOfBibitem
\bibitem[Kim \latin{et~al.}(2025)Kim, Han, Lee, Yoo, Park, Yu, Won, Lee, Ko,
  Noh, \latin{et~al.} others]{kim2025oxygen}
Kim,~W.; Han,~S.; Lee,~S.; Yoo,~J.; Park,~C.; Yu,~S.; Won,~D.; Lee,~E.;
  Ko,~K.-h.; Noh,~J.; others Oxygen-tuned aluminum-based halide solid
  electrolytes enabling low-voltage anode compatibility in all-solid-state
  batteries. \emph{Energy \& Environmental Science} \textbf{2025}, \emph{18},
  8039--8051\relax
\mciteBstWouldAddEndPuncttrue
\mciteSetBstMidEndSepPunct{\mcitedefaultmidpunct}
{\mcitedefaultendpunct}{\mcitedefaultseppunct}\relax
\EndOfBibitem
\bibitem[Huang \latin{et~al.}(2026)Huang, Zhang, Guo, Yang, Feng, Man, Zhang,
  Li, Du, Limphirat, \latin{et~al.} others]{huang2026amorphous}
Huang,~C.; Zhang,~G.; Guo,~H.; Yang,~L.; Feng,~Y.; Man,~Q.; Zhang,~Q.; Li,~C.;
  Du,~P.; Limphirat,~W.; others Amorphous aluminum-based oxychloride superionic
  conductors via cation--oxygen coupled modification for durable high-rate
  all-solid-state lithium batteries. \emph{Energy \& Environmental Science}
  \textbf{2026}, \emph{19}, 2084--2093\relax
\mciteBstWouldAddEndPuncttrue
\mciteSetBstMidEndSepPunct{\mcitedefaultmidpunct}
{\mcitedefaultendpunct}{\mcitedefaultseppunct}\relax
\EndOfBibitem
\bibitem[Palmer \latin{et~al.}(2026)Palmer, Singh, Merola, Pugazhendhi, Barai,
  You, Wang, Carlson, Wang, Zheng, \latin{et~al.} others]{palmer2026conformal}
Palmer,~M.; Singh,~V.~K.; Merola,~L.; Pugazhendhi,~K.~N.; Barai,~P.; You,~I.;
  Wang,~Y.; Carlson,~E.; Wang,~C.; Zheng,~H.; others Conformal
  Hetero-Electrolyte Interface between Soft Oxyhalides and Garnet Enables
  Low-Pressure Lithium Reservoir-Free Solid-State Batteries. \emph{Energy \&
  Environmental Science} \textbf{2026}, \relax
\mciteBstWouldAddEndPunctfalse
\mciteSetBstMidEndSepPunct{\mcitedefaultmidpunct}
{}{\mcitedefaultseppunct}\relax
\EndOfBibitem
\bibitem[Li \latin{et~al.}(2019)Li, Liang, Chen, Luo, Adair, Wang, Banis, Sham,
  Zhang, Zhao, \latin{et~al.} others]{li2019water}
Li,~X.; Liang,~J.; Chen,~N.; Luo,~J.; Adair,~K.~R.; Wang,~C.; Banis,~M.~N.;
  Sham,~T.-K.; Zhang,~L.; Zhao,~S.; others Water-mediated synthesis of a
  superionic halide solid electrolyte. \emph{Angewandte Chemie} \textbf{2019},
  \emph{131}, 16579--16584\relax
\mciteBstWouldAddEndPuncttrue
\mciteSetBstMidEndSepPunct{\mcitedefaultmidpunct}
{\mcitedefaultendpunct}{\mcitedefaultseppunct}\relax
\EndOfBibitem
\bibitem[Wang \latin{et~al.}(2021)Wang, Liang, Luo, Liu, Li, Zhao, Li, Huang,
  Zhao, Zhang, \latin{et~al.} others]{wang2021universal}
Wang,~C.; Liang,~J.; Luo,~J.; Liu,~J.; Li,~X.; Zhao,~F.; Li,~R.; Huang,~H.;
  Zhao,~S.; Zhang,~L.; others A universal wet-chemistry synthesis of
  solid-state halide electrolytes for all-solid-state lithium-metal batteries.
  \emph{Science advances} \textbf{2021}, \emph{7}, eabh1896\relax
\mciteBstWouldAddEndPuncttrue
\mciteSetBstMidEndSepPunct{\mcitedefaultmidpunct}
{\mcitedefaultendpunct}{\mcitedefaultseppunct}\relax
\EndOfBibitem
\bibitem[Gamo \latin{et~al.}(2025)Gamo, Kitta, and Takeichi]{gamo2025liquid}
Gamo,~H.; Kitta,~M.; Takeichi,~N. Liquid-Phase Synthesis of Halide Solid
  Electrolytes for All-Solid-State Batteries Using Organic Solvents.
  \emph{Energy \& Environmental Materials} \textbf{2025}, e70184\relax
\mciteBstWouldAddEndPuncttrue
\mciteSetBstMidEndSepPunct{\mcitedefaultmidpunct}
{\mcitedefaultendpunct}{\mcitedefaultseppunct}\relax
\EndOfBibitem
\bibitem[Kochetkov \latin{et~al.}(2022)Kochetkov, Zuo, Ruess, Singh, Zhou,
  Kaup, Janek, and Nazar]{kochetkov2022different}
Kochetkov,~I.; Zuo,~T.-T.; Ruess,~R.; Singh,~B.; Zhou,~L.; Kaup,~K.; Janek,~J.;
  Nazar,~L. Different interfacial reactivity of lithium metal chloride
  electrolytes with high voltage cathodes determines solid-state battery
  performance. \emph{Energy \& Environmental Science} \textbf{2022}, \emph{15},
  3933--3944\relax
\mciteBstWouldAddEndPuncttrue
\mciteSetBstMidEndSepPunct{\mcitedefaultmidpunct}
{\mcitedefaultendpunct}{\mcitedefaultseppunct}\relax
\EndOfBibitem
\bibitem[Zhang \latin{et~al.}(2025)Zhang, Zhao, Li, and Sun]{zhang2025solid}
Zhang,~S.; Zhao,~F.; Li,~L.; Sun,~X. Solid-state electrolytes expediting
  interface-compatible dual-conductive cathodes for all-solid-state batteries.
  \emph{Energy \& Environmental Science} \textbf{2025}, \emph{18},
  6530--6539\relax
\mciteBstWouldAddEndPuncttrue
\mciteSetBstMidEndSepPunct{\mcitedefaultmidpunct}
{\mcitedefaultendpunct}{\mcitedefaultseppunct}\relax
\EndOfBibitem
\bibitem[Tao \latin{et~al.}(2023)Tao, Zhong, Li, Wang, and
  Chang]{tao2023halide}
Tao,~B.; Zhong,~D.; Li,~H.; Wang,~G.; Chang,~H. Halide solid-state electrolytes
  for all-solid-state batteries: structural design, synthesis, environmental
  stability, interface optimization and challenges. \emph{Chemical science}
  \textbf{2023}, \emph{14}, 8693--8722\relax
\mciteBstWouldAddEndPuncttrue
\mciteSetBstMidEndSepPunct{\mcitedefaultmidpunct}
{\mcitedefaultendpunct}{\mcitedefaultseppunct}\relax
\EndOfBibitem
\bibitem[Tan \latin{et~al.}(2025)Tan, Beltran, Ke, Zhang, Choi, Zhou, Cho, Lee,
  and Su]{tan2025interfacial}
Tan,~Y.; Beltran,~M.; Ke,~J.; Zhang,~J.; Choi,~J.; Zhou,~Y.; Cho,~K.; Lee,~D.;
  Su,~L. Interfacial challenges of halide-based all-solid-state batteries.
  \emph{Advanced Energy Materials} \textbf{2025}, \emph{15}, 2403986\relax
\mciteBstWouldAddEndPuncttrue
\mciteSetBstMidEndSepPunct{\mcitedefaultmidpunct}
{\mcitedefaultendpunct}{\mcitedefaultseppunct}\relax
\EndOfBibitem
\bibitem[Li \latin{et~al.}(2024)Li, Wu, Cheng, Tan, Luo, Pan, Ma, Lu, Wen,
  Liang, \latin{et~al.} others]{li2024unraveling}
Li,~F.; Wu,~Y.-C.; Cheng,~X.-B.; Tan,~Y.; Luo,~J.-D.; Pan,~R.; Ma,~T.;
  Lu,~L.-L.; Wen,~X.; Liang,~Z.; others Unraveling the interfacial
  compatibility of ultrahigh nickel cathodes and chloride solid electrolyte for
  stable all-solid-state lithium batteries. \emph{Energy \& Environmental
  Science} \textbf{2024}, \emph{17}, 4187--4195\relax
\mciteBstWouldAddEndPuncttrue
\mciteSetBstMidEndSepPunct{\mcitedefaultmidpunct}
{\mcitedefaultendpunct}{\mcitedefaultseppunct}\relax
\EndOfBibitem
\bibitem[Fu \latin{et~al.}(2025)Fu, Wang, Wang, Reid, Liang, Luo, Kim, Zhao,
  Yang, Zhao, \latin{et~al.} others]{fu2025cost}
Fu,~J.; Wang,~C.; Wang,~S.; Reid,~J.~W.; Liang,~J.; Luo,~J.; Kim,~J.~T.;
  Zhao,~Y.; Yang,~X.; Zhao,~F.; others A cost-effective all-in-one halide
  material for all-solid-state batteries. \emph{Nature} \textbf{2025},
  \emph{643}, 111--118\relax
\mciteBstWouldAddEndPuncttrue
\mciteSetBstMidEndSepPunct{\mcitedefaultmidpunct}
{\mcitedefaultendpunct}{\mcitedefaultseppunct}\relax
\EndOfBibitem
\bibitem[Liang(1973)]{liangConductionCharacteristicsLithium1973}
Liang,~C.~C. Conduction {{Characteristics}} of the {{Lithium
  Iodide}}-{{Aluminum Oxide Solid Electrolytes}}. \emph{Journal of The
  Electrochemical Society} \textbf{1973}, \emph{120}, 1289\relax
\mciteBstWouldAddEndPuncttrue
\mciteSetBstMidEndSepPunct{\mcitedefaultmidpunct}
{\mcitedefaultendpunct}{\mcitedefaultseppunct}\relax
\EndOfBibitem
\bibitem[Sata \latin{et~al.}(2000)Sata, Eberman, Eberl, and
  Maier]{sataMesoscopicFastIon2000}
Sata,~N.; Eberman,~K.; Eberl,~K.; Maier,~J. Mesoscopic fast ion conduction in
  nanometre-scale planar heterostructures. \emph{Nature} \textbf{2000},
  \emph{408}, 946–949\relax
\mciteBstWouldAddEndPuncttrue
\mciteSetBstMidEndSepPunct{\mcitedefaultmidpunct}
{\mcitedefaultendpunct}{\mcitedefaultseppunct}\relax
\EndOfBibitem
\bibitem[Maier(1995)]{maierIonicConductionSpace1995}
Maier,~J. Ionic Conduction in Space Charge Regions. \emph{Progress in Solid
  State Chemistry} \textbf{1995}, \emph{23}, 171--263\relax
\mciteBstWouldAddEndPuncttrue
\mciteSetBstMidEndSepPunct{\mcitedefaultmidpunct}
{\mcitedefaultendpunct}{\mcitedefaultseppunct}\relax
\EndOfBibitem
\bibitem[Limon \latin{et~al.}(2026)Limon, Navid, Duffee, and
  Ahmad]{limonGrainBoundariesCeramic2026}
Limon,~M. S.~R.; Navid,~A.~F.; Duffee,~C.~W.; Ahmad,~Z. Grain {{Boundaries}} in
  {{Ceramic Solid-State Lithium Metal Batteries}}: {{A Review}}.
  \emph{Industrial \& Engineering Chemistry Research} \textbf{2026}, \emph{65},
  1--26\relax
\mciteBstWouldAddEndPuncttrue
\mciteSetBstMidEndSepPunct{\mcitedefaultmidpunct}
{\mcitedefaultendpunct}{\mcitedefaultseppunct}\relax
\EndOfBibitem
\bibitem[Ahmad(2025)]{ahmadUnifiedConsistentElectrical2025}
Ahmad,~Z. A Unified and Consistent Electrical Double Layer Model for the
  Treatment of Core and Space Charge Layers in Solid Electrolytes.
  \emph{Journal of Materials Chemistry A} \textbf{2025}, \emph{13},
  25599--25608\relax
\mciteBstWouldAddEndPuncttrue
\mciteSetBstMidEndSepPunct{\mcitedefaultmidpunct}
{\mcitedefaultendpunct}{\mcitedefaultseppunct}\relax
\EndOfBibitem
\bibitem[Wang \latin{et~al.}(2025)Wang, Limon, Zhou, Cho, Ahmad, and
  Su]{wang1+122025}
Wang,~B.; Limon,~M. S.~R.; Zhou,~Y.; Cho,~K.; Ahmad,~Z.; Su,~L. 1 +1 {$>$} 2
  {{Effect Induced}} by {{Space Charge}} in {{Solid Electrolytes}}. \emph{ACS
  Energy Letters} \textbf{2025}, \emph{10}, 1255--1257\relax
\mciteBstWouldAddEndPuncttrue
\mciteSetBstMidEndSepPunct{\mcitedefaultmidpunct}
{\mcitedefaultendpunct}{\mcitedefaultseppunct}\relax
\EndOfBibitem
\bibitem[Batzner \latin{et~al.}(2022)Batzner, Musaelian, Sun, Geiger, Mailoa,
  Kornbluth, Molinari, Smidt, and
  Kozinsky]{batznerE3EquivariantGraphNeural2022}
Batzner,~S.; Musaelian,~A.; Sun,~L.; Geiger,~M.; Mailoa,~J.~P.; Kornbluth,~M.;
  Molinari,~N.; Smidt,~T.~E.; Kozinsky,~B. E(3)-Equivariant Graph Neural
  Networks for Data-Efficient and Accurate Interatomic Potentials. \emph{Nature
  Communications} \textbf{2022}, \emph{13}, 2453\relax
\mciteBstWouldAddEndPuncttrue
\mciteSetBstMidEndSepPunct{\mcitedefaultmidpunct}
{\mcitedefaultendpunct}{\mcitedefaultseppunct}\relax
\EndOfBibitem
\bibitem[Maruf \latin{et~al.}(2025)Maruf, Kim, and
  Ahmad]{marufLearningLongRangeInteractions2025}
Maruf,~M.~U.; Kim,~S.; Ahmad,~Z. Learning {{Long-Range Interactions}} in
  {{Equivariant Machine Learning Interatomic Potentials}} via {{Electronic
  Degrees}} of {{Freedom}}. \emph{The Journal of Physical Chemistry Letters}
  \textbf{2025}, \emph{16}, 9078--9087\relax
\mciteBstWouldAddEndPuncttrue
\mciteSetBstMidEndSepPunct{\mcitedefaultmidpunct}
{\mcitedefaultendpunct}{\mcitedefaultseppunct}\relax
\EndOfBibitem
\bibitem[Deringer \latin{et~al.}(2019)Deringer, Caro, and
  Cs{\'a}nyi]{deringerMachineLearningInteratomic2019}
Deringer,~V.~L.; Caro,~M.~A.; Cs{\'a}nyi,~G. Machine {{Learning Interatomic
  Potentials}} as {{Emerging Tools}} for {{Materials Science}}. \emph{Advanced
  Materials} \textbf{2019}, \emph{31}, 1902765\relax
\mciteBstWouldAddEndPuncttrue
\mciteSetBstMidEndSepPunct{\mcitedefaultmidpunct}
{\mcitedefaultendpunct}{\mcitedefaultseppunct}\relax
\EndOfBibitem
\bibitem[Seth \latin{et~al.}(2025)Seth, Kulkarni, and
  Sai~Gautam]{seth2025investigating}
Seth,~A.; Kulkarni,~R.~P.; Sai~Gautam,~G. Investigating Ionic Diffusivity in
  Amorphous LiPON using Machine-Learned Interatomic Potentials. \emph{ACS
  Materials Au} \textbf{2025}, \emph{5}, 458--468\relax
\mciteBstWouldAddEndPuncttrue
\mciteSetBstMidEndSepPunct{\mcitedefaultmidpunct}
{\mcitedefaultendpunct}{\mcitedefaultseppunct}\relax
\EndOfBibitem
\bibitem[Pustorino \latin{et~al.}(2026)Pustorino, {Tapia-Aracayo}, Halbing,
  Brillson, Stewart, and Qi]{pustorinoMultiionDefectModel2026}
Pustorino,~G.; {Tapia-Aracayo},~L.~J.; Halbing,~D.; Brillson,~L.; Stewart,~D.;
  Qi,~Y. Multi-Ion Defect Model for Space-Charge Layer Formation at
  {{LiPON}}/{{LixV2O5}} Electrochemical Interfaces. \emph{npj Computational
  Materials} \textbf{2026}, DOI: 10.1038/s41524--026--02182--y\relax
\mciteBstWouldAddEndPuncttrue
\mciteSetBstMidEndSepPunct{\mcitedefaultmidpunct}
{\mcitedefaultendpunct}{\mcitedefaultseppunct}\relax
\EndOfBibitem
\bibitem[Limon and Ahmad(2024)Limon, and
  Ahmad]{limonHeterogeneityPointDefect2024c}
Limon,~M. S.~R.; Ahmad,~Z. Heterogeneity in {{Point Defect Distribution}} and
  {{Mobility}} in {{Solid Ion Conductors}}. \emph{ACS Applied Materials \&
  Interfaces} \textbf{2024}, \emph{16}, 50948--50960\relax
\mciteBstWouldAddEndPuncttrue
\mciteSetBstMidEndSepPunct{\mcitedefaultmidpunct}
{\mcitedefaultendpunct}{\mcitedefaultseppunct}\relax
\EndOfBibitem
\bibitem[Zhang \latin{et~al.}(2016)Zhang, Pan, Lu, Liu, Verbrugge, Sheldon,
  Cheng, Qi, and Xiao]{zhangSynergeticEffectsInorganic2016}
Zhang,~Q.; Pan,~J.; Lu,~P.; Liu,~Z.; Verbrugge,~M.~W.; Sheldon,~B.~W.;
  Cheng,~Y.-T.; Qi,~Y.; Xiao,~X. Synergetic {{Effects}} of {{Inorganic
  Components}} in {{Solid Electrolyte Interphase}} on {{High Cycle Efficiency}}
  of {{Lithium Ion Batteries}}. \emph{Nano Letters} \textbf{2016}, \emph{16},
  2011--2016\relax
\mciteBstWouldAddEndPuncttrue
\mciteSetBstMidEndSepPunct{\mcitedefaultmidpunct}
{\mcitedefaultendpunct}{\mcitedefaultseppunct}\relax
\EndOfBibitem
\bibitem[Pan \latin{et~al.}(2016)Pan, Zhang, Xiao, Cheng, and
  Qi]{panDesignNanostructuredHeterogeneous2016}
Pan,~J.; Zhang,~Q.; Xiao,~X.; Cheng,~Y.-T.; Qi,~Y. Design of {{Nanostructured
  Heterogeneous Solid Ionic Coatings}} through a {{Multiscale Defect Model}}.
  \emph{ACS Applied Materials \& Interfaces} \textbf{2016}, \emph{8},
  5687--5693\relax
\mciteBstWouldAddEndPuncttrue
\mciteSetBstMidEndSepPunct{\mcitedefaultmidpunct}
{\mcitedefaultendpunct}{\mcitedefaultseppunct}\relax
\EndOfBibitem
\bibitem[Ahmad \latin{et~al.}(2021)Ahmad, Venturi, Hafiz, and
  Viswanathan]{ahmadInterfacesSolidElectrolyte2021}
Ahmad,~Z.; Venturi,~V.; Hafiz,~H.; Viswanathan,~V. Interfaces in {{Solid
  Electrolyte Interphase}}: {{Implications}} for {{Lithium-Ion Batteries}}.
  \emph{The Journal of Physical Chemistry C} \textbf{2021}, \emph{125},
  11301--11309\relax
\mciteBstWouldAddEndPuncttrue
\mciteSetBstMidEndSepPunct{\mcitedefaultmidpunct}
{\mcitedefaultendpunct}{\mcitedefaultseppunct}\relax
\EndOfBibitem
\bibitem[Thompson \latin{et~al.}(2022)Thompson, Aktulga, Berger, Bolintineanu,
  Brown, Crozier, In't~Veld, Kohlmeyer, Moore, Nguyen, \latin{et~al.}
  others]{thompson2022lammps}
Thompson,~A.~P.; Aktulga,~H.~M.; Berger,~R.; Bolintineanu,~D.~S.; Brown,~W.~M.;
  Crozier,~P.~S.; In't~Veld,~P.~J.; Kohlmeyer,~A.; Moore,~S.~G.; Nguyen,~T.~D.;
  others LAMMPS-a flexible simulation tool for particle-based materials
  modeling at the atomic, meso, and continuum scales. \emph{Computer physics
  communications} \textbf{2022}, \emph{271}, 108171\relax
\mciteBstWouldAddEndPuncttrue
\mciteSetBstMidEndSepPunct{\mcitedefaultmidpunct}
{\mcitedefaultendpunct}{\mcitedefaultseppunct}\relax
\EndOfBibitem
\bibitem[Ahammed and Ertekin(2024)Ahammed, and
  Ertekin]{ahammedConfigurationalDisorderStrong2024}
Ahammed,~B.; Ertekin,~E. Configurational {{Disorder}}, {{Strong
  Anharmonicity}}, and {{Coupled Host Dynamics Lead}} to {{Superionic
  Transport}} in {{Li3YCl6}} ({{LYC}}). \emph{Advanced Materials}
  \textbf{2024}, \emph{36}, 2310537\relax
\mciteBstWouldAddEndPuncttrue
\mciteSetBstMidEndSepPunct{\mcitedefaultmidpunct}
{\mcitedefaultendpunct}{\mcitedefaultseppunct}\relax
\EndOfBibitem
\bibitem[Beltran \latin{et~al.}(2026)Beltran, Wang, Tan, Choi, Lee, and
  Su]{beltran2026mechanochemistry}
Beltran,~M.; Wang,~B.; Tan,~Y.; Choi,~J.; Lee,~D.; Su,~L.
  Mechanochemistry-Driven Optimization of Halide-Based Solid-State Electrolytes
  via Orthogonal Design of Experiments and Regression Modeling. \emph{ACS
  Materials Letters} \textbf{2026}, \relax
\mciteBstWouldAddEndPunctfalse
\mciteSetBstMidEndSepPunct{\mcitedefaultmidpunct}
{}{\mcitedefaultseppunct}\relax
\EndOfBibitem
\bibitem[{\'a}M~Photiadis \latin{et~al.}(1998){\'a}M~Photiadis, \latin{et~al.}
  others]{am1998vibrational}
{\'a}M~Photiadis,~G.; others Vibrational modes and structure of liquid and
  gaseous zirconium tetrachloride and of molten ZrCl 4--CsCl mixtures.
  \emph{Journal of the Chemical Society, Dalton Transactions} \textbf{1998},
  981--990\relax
\mciteBstWouldAddEndPuncttrue
\mciteSetBstMidEndSepPunct{\mcitedefaultmidpunct}
{\mcitedefaultendpunct}{\mcitedefaultseppunct}\relax
\EndOfBibitem
\bibitem[Brahmbhatt(2023)]{brahmbhatt2023synthesis}
Brahmbhatt,~T. The Synthesis and Optimization of Sulfide and Halide Solid
  Electrolytes for All Solid-State Batteries. \textbf{2023}, \relax
\mciteBstWouldAddEndPunctfalse
\mciteSetBstMidEndSepPunct{\mcitedefaultmidpunct}
{}{\mcitedefaultseppunct}\relax
\EndOfBibitem
\bibitem[Papatheodorou(1977)]{papatheodorou1977raman}
Papatheodorou,~G. Raman spectroscopic studies of yttrium (III) chloride--alkali
  metal chloride melts and of Cs2NaYCl6 and YCl3 solid compounds. \emph{The
  Journal of Chemical Physics} \textbf{1977}, \emph{66}, 2893--2900\relax
\mciteBstWouldAddEndPuncttrue
\mciteSetBstMidEndSepPunct{\mcitedefaultmidpunct}
{\mcitedefaultendpunct}{\mcitedefaultseppunct}\relax
\EndOfBibitem
\bibitem[Giannozzi \latin{et~al.}(2009)Giannozzi, Baroni, Bonini, Calandra,
  Car, Cavazzoni, Ceresoli, Chiarotti, Cococcioni, Dabo, Corso, de~Gironcoli,
  Fabris, Fratesi, Gebauer, Gerstmann, Gougoussis, Kokalj, Lazzeri,
  {Martin-Samos}, Marzari, Mauri, Mazzarello, Paolini, Pasquarello, Paulatto,
  Sbraccia, Scandolo, Sclauzero, Seitsonen, Smogunov, Umari, and
  Wentzcovitch]{giannozziQUANTUMESPRESSOModular2009}
Giannozzi,~P. \latin{et~al.}  {{QUANTUM ESPRESSO}}: A Modular and Open-Source
  Software Project for Quantum Simulations of Materials. \emph{J. Phys.:
  Condens. Matter} \textbf{2009}, \emph{21}, 395502\relax
\mciteBstWouldAddEndPuncttrue
\mciteSetBstMidEndSepPunct{\mcitedefaultmidpunct}
{\mcitedefaultendpunct}{\mcitedefaultseppunct}\relax
\EndOfBibitem
\bibitem[Giannozzi \latin{et~al.}(2017)Giannozzi, Andreussi, Brumme, Bunau,
  Nardelli, Calandra, Car, Cavazzoni, Ceresoli, Cococcioni, Colonna, Carnimeo,
  Corso, de~Gironcoli, Delugas, DiStasio, Ferretti, Floris, Fratesi, Fugallo,
  Gebauer, Gerstmann, Giustino, Gorni, Jia, Kawamura, Ko, Kokalj, K{\"u}{\c
  c}{\"u}kbenli, Lazzeri, Marsili, Marzari, Mauri, Nguyen, Nguyen,
  {Otero-de-la-Roza}, Paulatto, Ponc{\'e}, Rocca, Sabatini, Santra, Schlipf,
  Seitsonen, Smogunov, Timrov, Thonhauser, Umari, Vast, Wu, and
  Baroni]{giannozziAdvancedCapabilitiesMaterials2017}
Giannozzi,~P. \latin{et~al.}  Advanced Capabilities for Materials Modelling
  with {{Quantum ESPRESSO}}. \emph{Journal of Physics: Condensed Matter}
  \textbf{2017}, \emph{29}, 465901\relax
\mciteBstWouldAddEndPuncttrue
\mciteSetBstMidEndSepPunct{\mcitedefaultmidpunct}
{\mcitedefaultendpunct}{\mcitedefaultseppunct}\relax
\EndOfBibitem
\bibitem[Perdew \latin{et~al.}(1996)Perdew, Burke, and
  Ernzerhof]{perdewGeneralizedGradientApproximation1996}
Perdew,~J.~P.; Burke,~K.; Ernzerhof,~M. Generalized {{Gradient Approximation
  Made Simple}}. \emph{Physical Review Letters} \textbf{1996}, \emph{77},
  3865--3868\relax
\mciteBstWouldAddEndPuncttrue
\mciteSetBstMidEndSepPunct{\mcitedefaultmidpunct}
{\mcitedefaultendpunct}{\mcitedefaultseppunct}\relax
\EndOfBibitem
\bibitem[Garrity \latin{et~al.}(2014)Garrity, Bennett, Rabe, and
  Vanderbilt]{garrityPseudopotentialsHighthroughputDFT2014}
Garrity,~K.~F.; Bennett,~J.~W.; Rabe,~K.~M.; Vanderbilt,~D. Pseudopotentials
  for High-Throughput {{DFT}} Calculations. \emph{Computational Materials
  Science} \textbf{2014}, \emph{81}, 446--452\relax
\mciteBstWouldAddEndPuncttrue
\mciteSetBstMidEndSepPunct{\mcitedefaultmidpunct}
{\mcitedefaultendpunct}{\mcitedefaultseppunct}\relax
\EndOfBibitem
\bibitem[Grimme \latin{et~al.}(2010)Grimme, Antony, Ehrlich, and
  Krieg]{Grimme2010}
Grimme,~S.; Antony,~J.; Ehrlich,~S.; Krieg,~H. A consistent and accurate ab
  initio parametrization of density functional dispersion correction ({DFT}-D)
  for the 94 elements H-Pu. \emph{The Journal of Chemical Physics}
  \textbf{2010}, \emph{132}, 154104\relax
\mciteBstWouldAddEndPuncttrue
\mciteSetBstMidEndSepPunct{\mcitedefaultmidpunct}
{\mcitedefaultendpunct}{\mcitedefaultseppunct}\relax
\EndOfBibitem
\bibitem[Ong \latin{et~al.}(2013)Ong, Richards, Jain, Hautier, Kocher, Cholia,
  Gunter, Chevrier, Persson, and Ceder]{Ong2013pymatgen}
Ong,~S.~P.; Richards,~W.~D.; Jain,~A.; Hautier,~G.; Kocher,~M.; Cholia,~S.;
  Gunter,~D.; Chevrier,~V.~L.; Persson,~K.~A.; Ceder,~G. Python {{Materials
  Genomics}} (Pymatgen): {{A}} Robust, Open-Source Python Library for Materials
  Analysis. \emph{Computational Materials Science} \textbf{2013}, \emph{68},
  314--319\relax
\mciteBstWouldAddEndPuncttrue
\mciteSetBstMidEndSepPunct{\mcitedefaultmidpunct}
{\mcitedefaultendpunct}{\mcitedefaultseppunct}\relax
\EndOfBibitem
\bibitem[Larsen \latin{et~al.}(2017)Larsen, Mortensen, Blomqvist, Castelli,
  Christensen, Du{\l}ak, Friis, Groves, Hammer, Hargus, Hermes, Jennings,
  Jensen, Kermode, Kitchin, Kolsbjerg, Kubal, Kaasbjerg, Lysgaard, Maronsson,
  Maxson, Olsen, Pastewka, Peterson, Rostgaard, Schi{\o}tz, Sch\"{u}tt,
  Strange, Thygesen, Vegge, Vilhelmsen, Walter, Zeng, and
  Jacobsen]{HjorthLarsen2017atomic}
Larsen,~A.~H. \latin{et~al.}  The atomic simulation environment{\textemdash}a
  Python library for working with atoms. \emph{J. Phys.: Condens. Matter}
  \textbf{2017}, \emph{29}, 273002\relax
\mciteBstWouldAddEndPuncttrue
\mciteSetBstMidEndSepPunct{\mcitedefaultmidpunct}
{\mcitedefaultendpunct}{\mcitedefaultseppunct}\relax
\EndOfBibitem
\bibitem[Momma and Izumi(2011)Momma, and Izumi]{Momma2011vesta}
Momma,~K.; Izumi,~F. {VESTA} 3 for three-dimensional visualization of crystal,
  volumetric and morphology data. \emph{J. Appl. Crystallogr.} \textbf{2011},
  \emph{44}, 1272--1276\relax
\mciteBstWouldAddEndPuncttrue
\mciteSetBstMidEndSepPunct{\mcitedefaultmidpunct}
{\mcitedefaultendpunct}{\mcitedefaultseppunct}\relax
\EndOfBibitem
\bibitem[Stegmaier \latin{et~al.}(2017)Stegmaier, Voss, Reuter, and
  Luntz]{stegmaierLiDefectsSolidState2017}
Stegmaier,~S.; Voss,~J.; Reuter,~K.; Luntz,~A.~C. Li+ {{Defects}} in a
  {{Solid-State Li Ion Battery}}: {{Theoretical Insights}} with a {{Li3OCl
  Electrolyte}}. \emph{Chemistry of Materials} \textbf{2017}, \emph{29},
  4330--4340\relax
\mciteBstWouldAddEndPuncttrue
\mciteSetBstMidEndSepPunct{\mcitedefaultmidpunct}
{\mcitedefaultendpunct}{\mcitedefaultseppunct}\relax
\EndOfBibitem
\bibitem[Freysoldt \latin{et~al.}(2009)Freysoldt, Neugebauer, and {Van de
  Walle}]{freysoldtFullyInitioFiniteSize2009}
Freysoldt,~C.; Neugebauer,~J.; {Van de Walle},~C.~G. Fully {{{\emph{Ab
  Initio}}}} {{Finite-Size Corrections}} for {{Charged-Defect Supercell
  Calculations}}. \emph{Physical Review Letters} \textbf{2009}, \emph{102},
  016402\relax
\mciteBstWouldAddEndPuncttrue
\mciteSetBstMidEndSepPunct{\mcitedefaultmidpunct}
{\mcitedefaultendpunct}{\mcitedefaultseppunct}\relax
\EndOfBibitem
\bibitem[Kim \latin{et~al.}(2020)Kim, Hood, Park, Whalley, and
  Walsh]{kim2020quick}
Kim,~S.; Hood,~S.~N.; Park,~J.-S.; Whalley,~L.~D.; Walsh,~A. Quick-start guide
  for first-principles modelling of point defects in crystalline materials.
  \emph{Journal of Physics: Energy} \textbf{2020}, \emph{2}, 036001\relax
\mciteBstWouldAddEndPuncttrue
\mciteSetBstMidEndSepPunct{\mcitedefaultmidpunct}
{\mcitedefaultendpunct}{\mcitedefaultseppunct}\relax
\EndOfBibitem
\bibitem[Freysoldt and Neugebauer(2018)Freysoldt, and
  Neugebauer]{freysoldt2018first}
Freysoldt,~C.; Neugebauer,~J. First-principles calculations for charged defects
  at surfaces, interfaces, and two-dimensional materials in the presence of
  electric fields. \emph{Physical Review B} \textbf{2018}, \emph{97},
  205425\relax
\mciteBstWouldAddEndPuncttrue
\mciteSetBstMidEndSepPunct{\mcitedefaultmidpunct}
{\mcitedefaultendpunct}{\mcitedefaultseppunct}\relax
\EndOfBibitem
\bibitem[Neese(2012)]{ORCA}
Neese,~F. The ORCA program system. \emph{WIRES Comput. Molec. Sci.}
  \textbf{2012}, \emph{2}, 73--78\relax
\mciteBstWouldAddEndPuncttrue
\mciteSetBstMidEndSepPunct{\mcitedefaultmidpunct}
{\mcitedefaultendpunct}{\mcitedefaultseppunct}\relax
\EndOfBibitem
\bibitem[Neese(2022)]{ORCA5}
Neese,~F. Software update: the ORCA program system, version 5.0. \emph{WIRES
  Comput. Molec. Sci.} \textbf{2022}, \emph{12}, e1606\relax
\mciteBstWouldAddEndPuncttrue
\mciteSetBstMidEndSepPunct{\mcitedefaultmidpunct}
{\mcitedefaultendpunct}{\mcitedefaultseppunct}\relax
\EndOfBibitem
\bibitem[Adamo and Barone(1999)Adamo, and Barone]{Adamo1999}
Adamo,~C.; Barone,~V. Toward reliable density functional methods without
  adjustable parameters: The PBE0 model. \emph{The Journal of Chemical Physics}
  \textbf{1999}, \emph{110}, 6158–6170\relax
\mciteBstWouldAddEndPuncttrue
\mciteSetBstMidEndSepPunct{\mcitedefaultmidpunct}
{\mcitedefaultendpunct}{\mcitedefaultseppunct}\relax
\EndOfBibitem
\bibitem[Grimme \latin{et~al.}(2011)Grimme, Ehrlich, and Goerigk]{Grimme2011}
Grimme,~S.; Ehrlich,~S.; Goerigk,~L. Effect of the damping function in
  dispersion corrected density functional theory. \emph{Journal of
  Computational Chemistry} \textbf{2011}, \emph{32}, 1456–1465\relax
\mciteBstWouldAddEndPuncttrue
\mciteSetBstMidEndSepPunct{\mcitedefaultmidpunct}
{\mcitedefaultendpunct}{\mcitedefaultseppunct}\relax
\EndOfBibitem
\bibitem[Weigend and Ahlrichs(2005)Weigend, and Ahlrichs]{Weigend2005}
Weigend,~F.; Ahlrichs,~R. Balanced basis sets of split valence, triple zeta
  valence and quadruple zeta valence quality for H to Rn: Design and assessment
  of accuracy. \emph{Physical Chemistry Chemical Physics} \textbf{2005},
  \emph{7}, 3297\relax
\mciteBstWouldAddEndPuncttrue
\mciteSetBstMidEndSepPunct{\mcitedefaultmidpunct}
{\mcitedefaultendpunct}{\mcitedefaultseppunct}\relax
\EndOfBibitem
\end{mcitethebibliography}
\clearpage


\section*{Supplementary material}

# Supplementary Note 1: Interface generation

The crystal structures of LYC and LZC were taken from Asano et al.[1] and Kwak et al.[2], respectively. The relaxed bulk lattice parameters are $a = b = 22.3938$ Å and $c = 12.06224$ Å for LYC, and $a = 12.77244$ Å, $b = 22.10318$ Å, and $c = 12.59764$ Å for LZC. The generated 2 by 2 by 2 LYC supercell was slightly off stoichiometry, $Li_3Y_{1.06}Cl_{6.18}$ ($Li_{68}Y_{24}Cl_{140}$) in order to satisfy the fractional occupancies from Asano et al.[1] and the charge neutrality constraint. The generated 2 by 2 by 2 LZC supercell was stoichiometric.

The coherent interface was constructed using matched in-plane lattice constants of $a =$ 12.41700 Å and $b = 22.24850$ Å. Relative to the relaxed bulk references, this imposes biaxial in-plane strains of $\varepsilon_a = -2.78\%$ and $\varepsilon_b = +0.66\%$ for LZC (matched in the $ab$ plane), and $\varepsilon_c = +2.94\%$ and $\varepsilon_a = -0.65\%$ for LYC (matched in the $ca$ plane).

Table S1: Defect formation energies of selected charge-compensating Li defect pairs in LYC and LZC, evaluated at the charge-neutrality Fermi level (CNFL), together with their summed formation energies. Subscripts $fZ/nZ$ and $fY/nY$ denote defect sites farther from or nearer to the nearest Zr or Y atom, respectively.

| LYC | | LZC | | Sum (eV) |
|---|---|---|---|---|
| Defect | DFE (eV) | Defect | DFE (eV) | |
| $(\mathrm{Li}_i^\bullet)_{fY}$ | $-1.73$ | $(V_{\mathrm{Li}}^{'})_{fZ}$ | 1.57 | $-0.16$ |
| $(\mathrm{Li}_i^\bullet)_{nY}$ | $-1.55$ | $(V_{\mathrm{Li}}^{'})_{nZ}$ | 1.97 | 0.42 |
| $(\mathrm{Li}_i^\bullet)_{fY}$ | $-1.73$ | $(V_{\mathrm{Li}}^{'})_{nZ}$ | 1.97 | 0.24 |
| $(\mathrm{Li}_i^\bullet)_{nY}$ | $-1.55$ | $(V_{\mathrm{Li}}^{'})_{fZ}$ | 1.57 | 0.02 |
| $(V_{\mathrm{Li}}^{'})_{nY}$ | 1.11 | $(\mathrm{Li}_i^\bullet)_{fZ}$ | $-2.23$ | $-1.12$ |
| $(V_{\mathrm{Li}}^{'})_{fY}$ | 1.79 | $(\mathrm{Li}_i^\bullet)_{nZ}$ | $-1.40$ | 0.39 |
| $(V_{\mathrm{Li}}^{'})_{nY}$ | 1.11 | $(\mathrm{Li}_i^\bullet)_{nZ}$ | $-1.40$ | $-0.29$ |
| $(V_{\mathrm{Li}}^{'})_{fY}$ | 1.79 | $(\mathrm{Li}_i^\bullet)_{fZ}$ | $-2.23$ | $-0.44$ |

Table S2: Defect formation energies of selected charge-compensating Cl defect pairs in LYC and LZC, evaluated at their respective charge-neutrality Fermi levels (CNFLs), together with their summed formation energies. Subscripts $fZ/nZ$ and $fY/nY$ denote defect sites farther from or nearer to the nearest Zr or Y atom, respectively.

| LYC | | LZC | | |
|---|---|---|---|---|
| Defect | DFE (eV) | Defect | DFE (eV) | Sum (eV) |
| $(Cl_i')_{fY}$ | 1.04 | $(V_{\mathrm{Cl}}^\bullet)_{fZ}$ | -2.03 | -0.99 |
| $(Cl_i')_{nY}$ | 2.00 | $(V_{\mathrm{Cl}}^\bullet)_{nZ}$ | -2.03 | -0.03 |
| $(Cl_i')_{fY}$ | 1.04 | $(V_{\mathrm{Cl}}^\bullet)_{nZ}$ | -2.03 | **-0.99** |
| $(Cl_i')_{nY}$ | 2.00 | $(V_{\mathrm{Cl}}^\bullet)_{fZ}$ | -2.03 | -0.03 |
| $(V_{\mathrm{Cl}}^\bullet)_{nY}$ | -0.37 | $(Cl_i')_{fZ}$ | 3.36 | 2.99 |
| $(V_{\mathrm{Cl}}^\bullet)_{fY}$ | -1.28 | $(Cl_i')_{nZ}$ | 0.30 | **-0.98** |
| $(V_{\mathrm{Cl}}^\bullet)_{nY}$ | -0.37 | $(Cl_i')_{nZ}$ | 0.30 | -0.07 |
| $(V_{\mathrm{Cl}}^\bullet)_{fY}$ | -1.28 | $(Cl_i')_{fZ}$ | 3.36 | 2.08 |

(a) LZC dielectric tensor

```
Electronic dielectric permittivity tensor (F/m units)
        3.356177    0.009827    0.021451
        0.009828    3.211337   -0.019810
        0.021451   -0.019811    3.299997

 ... with zone-center polar mode contributions
        8.703713    0.796154    0.937427
        0.796155    8.524468   -3.257547
        0.937427   -3.257548   15.892823
```

(b) LYC dielectric tensor

```
Electronic dielectric permittivity tensor (F/m units)
        2.960376   -0.003051    0.011147
       -0.003051    2.967923   -0.007681
        0.011148   -0.007681    2.909822

 ... with zone-center polar mode contributions
        7.182208   -0.109407   -0.117258
       -0.109407    6.368085   -0.019591
       -0.117258   -0.019591    7.013946
```

Figure S1: Electronic dielectric permittivity tensors and corresponding total dielectric tensors including zone-center polar-mode contributions for (a) $Li_2ZrCl_6$ (LZC) and (b) $Li_3YCl_6$ (LYC). The upper $3\times3$ matrix in each panel gives the electronic contribution, while the lower matrix includes the additional ionic contribution from zone-center polar phonon modes.

**LZC TDOS**

Highest occupied, Lowest unoccupied level (ev): 2.9653 4.8155

Bandgap is 1.8502 eV

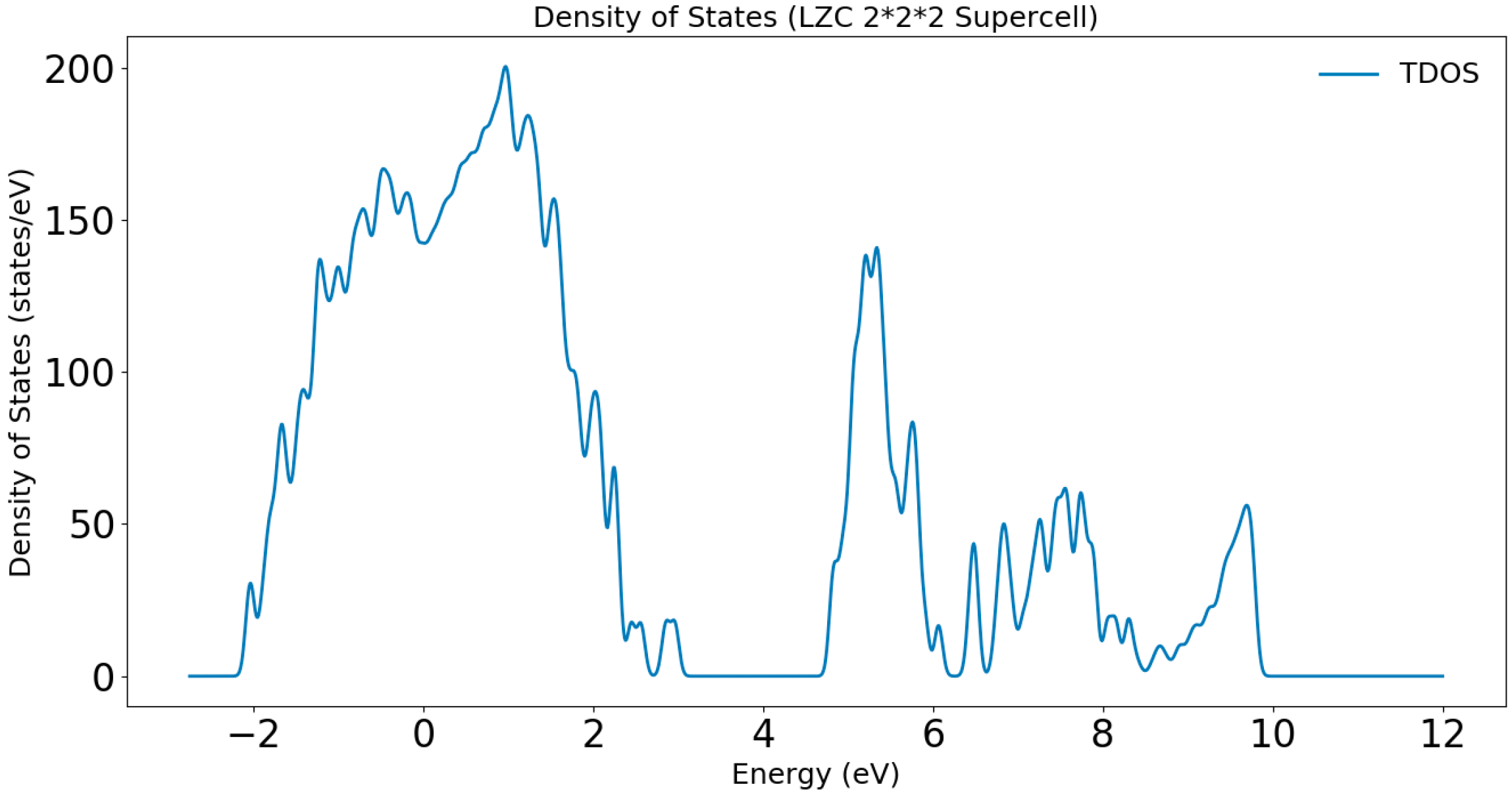


**LYC TDOS**

Highest occupied, Lowest unoccupied level (ev): 2.2763 5.2090

Bandgap is 2.9327 eV

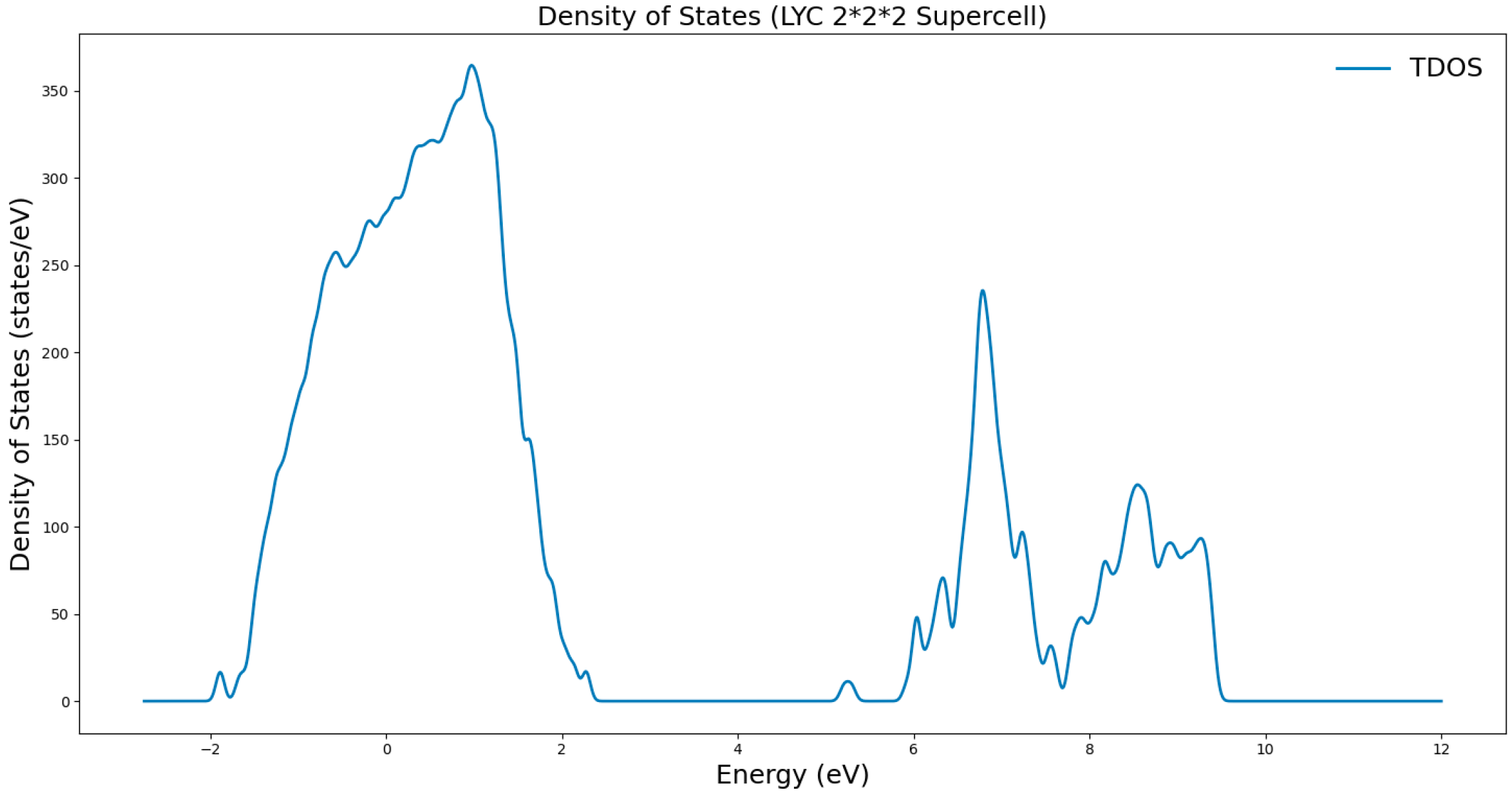


Figure S2: Total density of states (TDOS) for $Li_2ZrCl_6$ (LZC) and $Li_3YCl_6$ (LYC) $2 \times 2 \times 2$ supercells. The calculated highest occupied and lowest unoccupied energy levels are 2.9653 and 4.8155 eV for LZC and 2.2763 and 5.2090 eV for LYC, corresponding to band gaps of 1.8502 and 2.9327 eV, respectively. These are smaller than experimental values due to systematic underestimation by PBE functional.

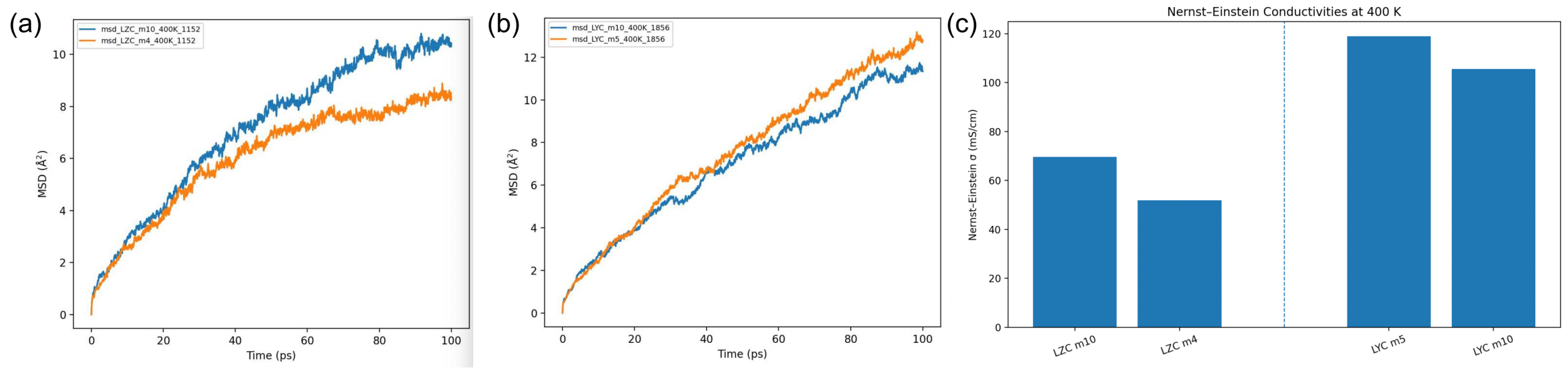


Figure S3: Comparison of Li-ion transport in independently generated LZC and LYC $2\times2\times2$ supercells. For each material, ten supercell configurations were generated by selecting Li occupancies according to the crystallographic site-occupancy information, followed by structural relaxation; the lowest-energy configuration was selected for the main calculations. To assess the sensitivity of Li transport to the specific Li distribution, additional low-energy configurations were also subjected to molecular dynamics simulations at 400 K. (a) Li mean-squared displacement (MSD) for two LZC configurations and (b) Li MSD for two LYC configurations. (c) Corresponding Nernst–Einstein Li-ion conductivities at 400 K. The comparable MSD trends and conductivities indicate that the predicted Li-ion transport is not strongly dependent on the particular Li arrangement among the low-energy supercell configurations considered.

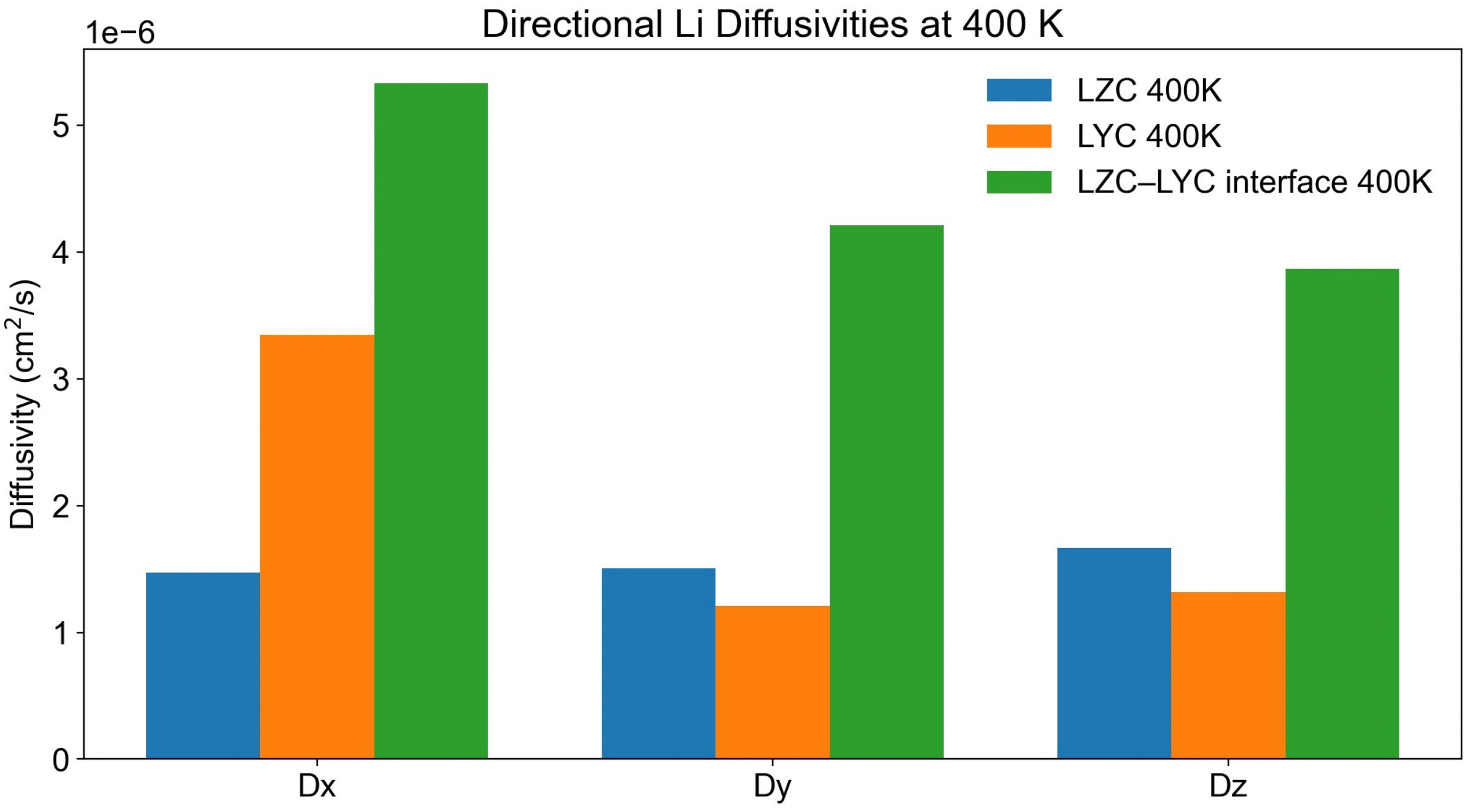


Figure S4: Direction-resolved Li-ion diffusivities at 400 K for LZC, LYC, and the LZC|LYC interface superlattice. The diffusivities along the Cartesian $x$-, $y$-, and $z$-directions are compared for all three systems. The interface exhibits higher Li diffusivity in all directions than the corresponding bulk LZC and LYC structures, with the largest enhancement along the $x$-direction.

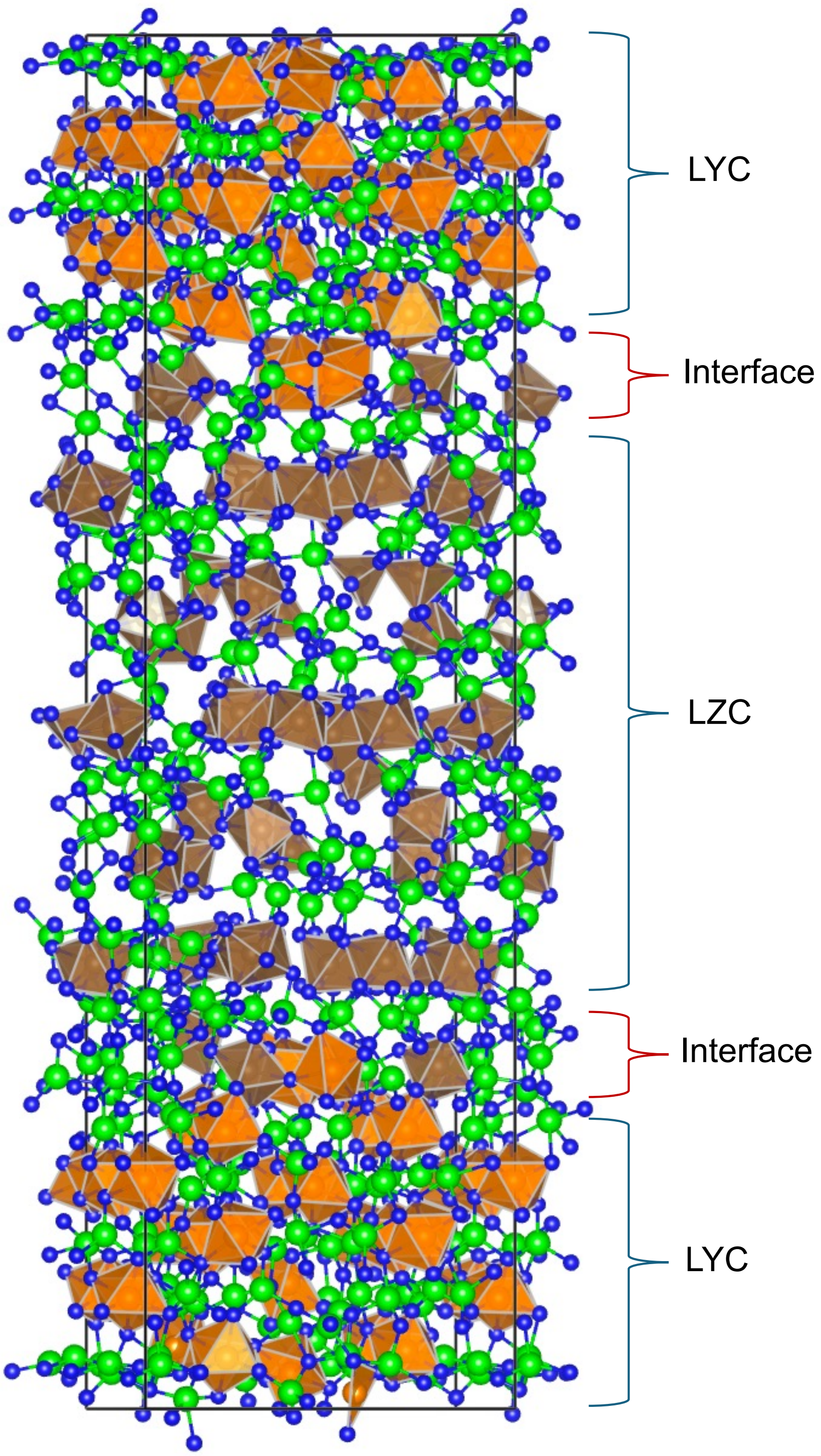


Figure S5: Periodic LYC|LZC superlattice unit cell used as the building block for the NequIP-based molecular dynamics models. The superlattice was constructed by stacking three $2\times2\times2$ LZC supercells between two $2\times2\times2$ LYC supercells along the interface normal, yielding an 896-atom cell containing two equivalent LYC|LZC interfaces separated by bulk-like regions of each phase. For molecular dynamics simulations, this unit cell was replicated $2\times2\times1$ in $(x, y, z)$ to form the larger periodic superlattice model used in the transport calculations.

# Supplementary Note 2: Effects of defects on bulk transport

Besides pristine structures, we also performed MD simulations on defective structures to understand their impact on Li ion transport. The point defects introduced in the MD models were chosen based on the DFE trends in Fig. 3: for both LZC and LYC, $\text{Li}_i$ and $V_{\text{Cl}}$ are the lowest-energy donor-like defects and become favorable near the CNFL/CBM, whereas $V_{\text{Li}}$ is stabilized closer to the VBM and remains comparatively higher in energy at the CNFL. Figure S6 shows Arrhenius plots of Nernst–Einstein conductivity, where all systems show approximately activated behavior over the simulated temperature range. In LZC, $V_{\text{Li}}$ and $V_{\text{Cl}}$ modestly lower the apparent activation barrier relative to pristine LZC, while $\text{Li}_i$ yields a similar barrier but tends to increase the conductivity at higher temperatures, due to an increased population of mobile Li carriers. In LYC, each defect produces a slightly reduced activation energy compared to pristine LYC, with the donor-like defects ($\text{Li}_i$ and $V_{\text{Cl}}$) giving the most consistent enhancement in conductivity across temperatures. Overall, the transport trends align with the defect thermodynamics: the low-DFE donor-like defects identified in Fig. 3 are also the most effective at improving bulk Li ion transport.

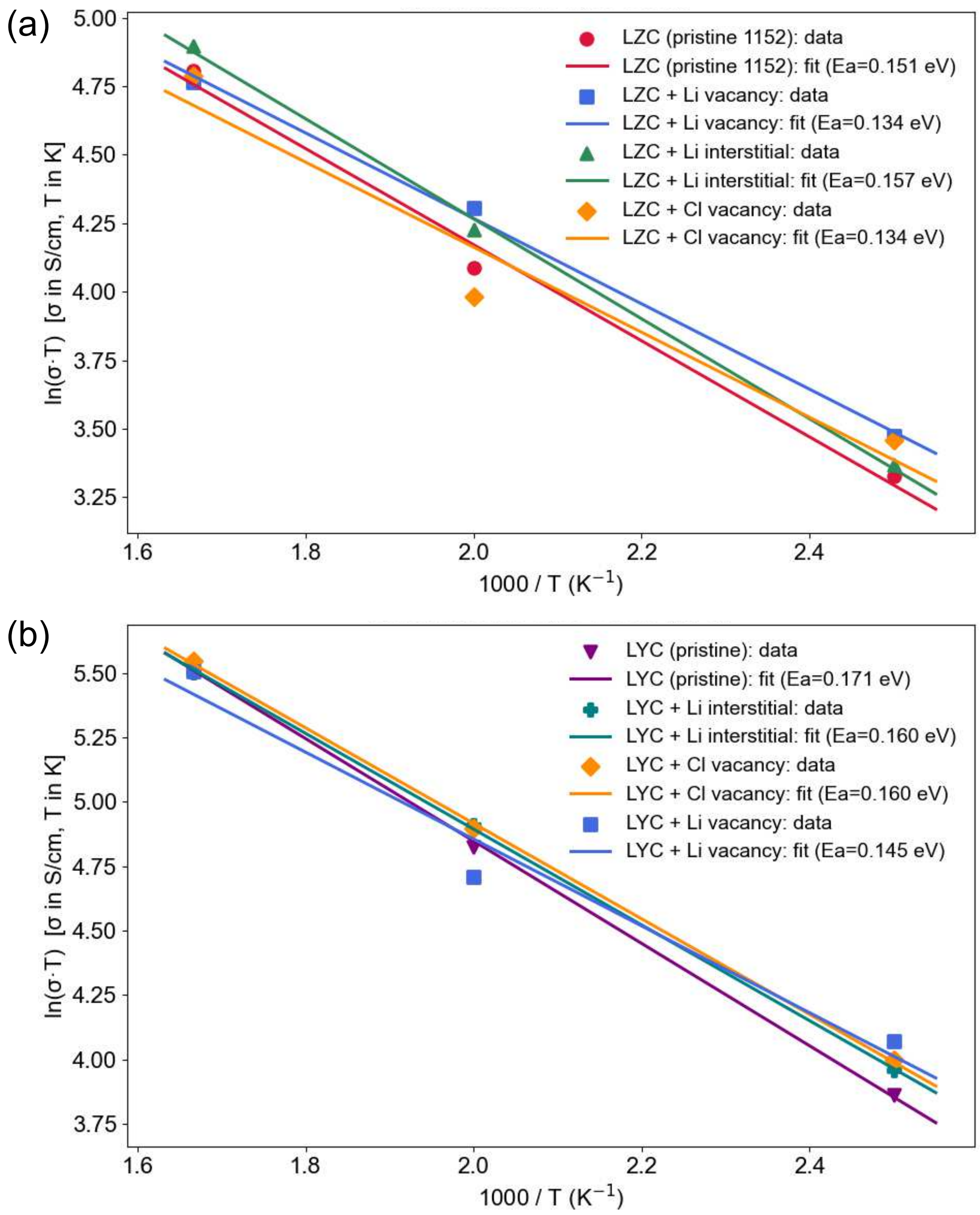


Figure S6: Arrhenius analysis of Li-ion transport in the bulk halide solid electrolytes from MD simulations for (a) LZC and (b) LYC. Shown are $\ln(\sigma T)$ (with $\sigma$ the Nernst–Einstein ionic conductivity) as a function of $1000/T$ for (a) bulk LZC (1152-atom $4 \times 4 \times 4$ supercell) and (b) bulk LYC (1856-atom $4 \times 4 \times 4$ supercell), comparing pristine structures and representative point-defect variants ($V_{\mathrm{Li}}$, $\mathrm{Li}_i$, and $V_{\mathrm{Cl}}$, as labeled). Symbols denote MD data at each temperature and solid lines are linear fits used to extract the effective activation energies $E_{\mathrm{a}}$ (reported in the legend). The defect chemistries were selected based on the DFT defect-formation-energy trends: $\mathrm{Li}_i$ and $V_{\mathrm{Cl}}$ are the lowest-energy donor-like defects and become favorable near the CNFL/CBM, whereas $V_{\mathrm{Li}}$ is stabilized closer to the VBM and remains higher in energy at the CNFL, enabling a transport-level comparison of thermodynamically likely versus less favorable native defects. The larger replicated cells preserve the local environments captured by the DFT-trained potential while reducing finite-size effects and improving statistical sampling for reliable diffusivity and conductivity estimates.

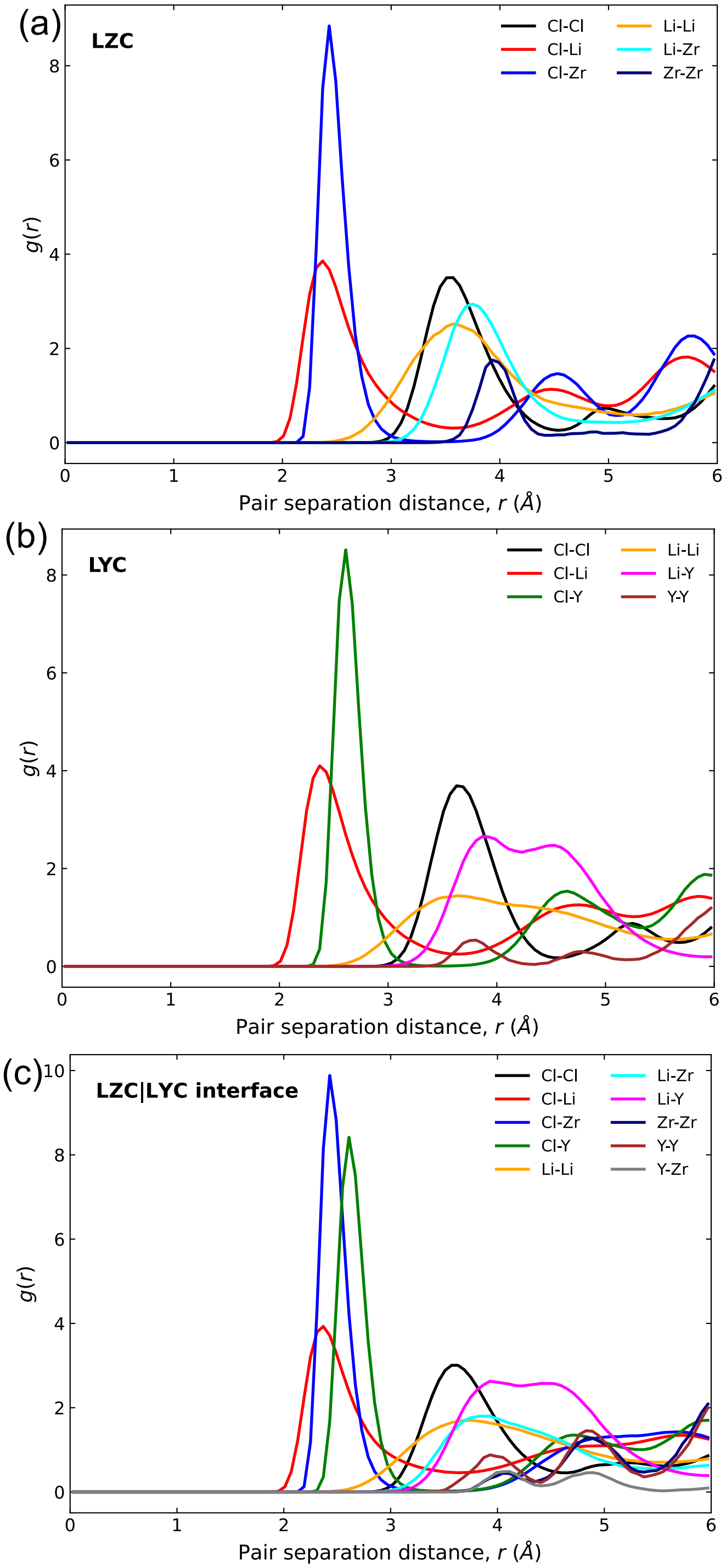


Figure S7: Partial radial distribution functions (RDFs), $g(r)$, calculated from the final 10 ps (90–100 ps) of the MD trajectories for (a) bulk LZC, (b) bulk LYC, and (c) the LZC|LYC interface superlattice. The RDFs were averaged over all sampled configurations within this time interval. The pair-resolved distributions show the characteristic short-range coordination environments of Li, Cl, Zr, and Y in the bulk phases and enable direct comparison with the corresponding local structural environments in the LZC|LYC interface.


\end{document}